%% file: optica_revised.tex
\documentclass[9pt,twocolumn,twoside]{opticajnl}
\journal{opticajournal} 

\definecolor{color2}{RGB}{59,90,198} 

\setboolean{shortarticle}{false}

\usepackage{lineno}

\graphicspath{{paper_figs/}}

\usepackage{bm}

\usepackage{hyperref}
 \hypersetup{
     colorlinks=true,
     linkcolor=blue,
     filecolor=blue,
     citecolor = magenta,      
     urlcolor=red,
     }

\usepackage{graphicx} 

\usepackage{amsthm,amsmath,amssymb,amsfonts,bbm}
\usepackage{braket}
\usepackage{bbold}
\usepackage{float}
\usepackage[capitalise]{cleveref}

\newcommand{\hc}{\text{H.c.}}

\definecolor{orange}{RGB}{200,100,0}

\usepackage{standalone}
\usepackage{tikz}
\usetikzlibrary{automata, positioning,calc, quotes}

\title{Pulsed two-photon scattering from a single atom in a waveguide with delay-modified temporal correlations}

\author[1,*]{Matthew Kozma}
\author[1]{Sofia Arranz Regidor}
\author[1]{Stephen Hughes}

\affil[1]{Department of Physics,
Engineering Physics and Astronomy, Queen's University, Kingston, Ontario, Canada, K7L 3N6}

\affil[*]{24nkr1@queensu.ca}

\begin{abstract} 
Quantum nonlinearity is an essential ingredient for many quantum technologies, but is often too weak to be exploited at the few-photon level. However, few photon fields interacting strongly with single quantum emitters in a waveguide environment can impact a significant nonlinear response, opening up a wide range of photon-photon correlations at a very fundamental level. 
Considering a waveguide-QED system containing a single atom (treated as a two-level system) chirally coupled to a waveguide, we theoretically investigate two-photon nonlinearities with delay-controlled temporal correlations. We use both matrix product states (MPS) and a frequency-dependent scattering theory approach to analyze the exact population dynamics, as well as the first-order and second-order photon correlation functions in transmission of the system, when pumped by a two-photon Fock-state pulse with a bimodal temporal pulse envelope. The two-photon Fock-state pulses are considered to be either two single photons localized to each peak of the pulse, or both photons delocalized between the two peaks. We consider the regimes of a short, moderate, and (relatively) long distance between the two pulse peaks, comparing the important differences in the temporal correlations with the two types of two-photon pulses. We demonstrate the strikingly different nonlinear features and quantum correlations that occur for uncorrelated and correlated two-photon pairs in experimentally accessible regimes.

\end{abstract}

\setboolean{displaycopyright}{false} 

\begin{document}

\maketitle

\section{Introduction}
\label{sec:intro}

The study of quantized light (photons) is 
fundamentally interesting and 
important in the development of many quantum technology applications in computing, communications, and sensing. From a practical perspective, waveguide quantum electrodynamics (waveguide-QED) confines quantized light to waveguide channels as one-dimensional propagating modes, which strongly couples photons and matter and acts as a desirable interface for studying the propagation of information~\cite{PhysRevA.95.033818,doi:10.1021/acs.nanolett.0c00607,PhysRevLett.104.023602,
PhysRevA.104.L031701,Whalen2017,PhysRevA.104.053701,
Liu:22,PhysRevResearch.2.043184,PhysRevA.105.033705,
PhysRevB.81.155117,Chang_2012,
PhysRevA.65.033832,Fang2018,Shen:05,Calajo2019,
PhysRevA.98.043816,PhysRevLett.111.243602,
Droenner2019,Cheng2017,PhysRevA.83.063828,
Zheng2010,RevModPhys.95.015002,
PhysRevA.79.023837,PhysRevA.76.062709,PhysRevLett.116.093601,
PhysRevLett.98.153003,PhysRevA.101.023807,PhysRevA.106.013714,Chen_2011}.

While desired transformations of single photons can be implemented using components from linear optics, constructing the necessary two-photon optical gates for many applications requires the presence of nonlinear interactions. Due to the typical weakness of classical nonlinearities, many groups are now considering the use of quantum nonlinearities in the form of quantized matter systems strongly coupled to the propagating light field. The transition structure of these matter systems may take on many different forms depending on their implementation, though a common and simple example of these structures is a quantum dot~\cite{PhysRevLett.101.113903,le_jeannic_dynamical_2022} or flux qubit~\cite{Astafiev2010,PhysRevLett.108.263601,vanLoo2013}, which may be approximated as a two-level system (TLS).

The basic understanding of single photon dynamics coupled to a TLS in waveguide-QED has been well studied, both in the continuous wave (CW) case---in which perfect reflection occurs on resonance (neglecting additional dissipation, apart from radiative decay to the 
waveguide)---and in the case of pulsed light. The linear optical nature of the interactions with the TLS within the single photon subspace makes studying this system relatively easy. Even so, interesting effects can be observed in the case of the single photon interaction, with the potential for significant population transfer to the TLS and even complete inversion provided the photon has the proper temporal/frequency pulse envelope~\cite{PhysRevA.82.033804,sofia2025,Chen_2011}.

In contrast to the single photon (linear) case, two-photon interactions with a TLS can yield nonlinear phenomena due to saturation effects of the TLS upon exit from the single quanta subspace. Furthermore, the quantum dynamics are complicated by the elastic scattering component from the TLS inducing correlations between the scattered photons that would not be present in the input pulse. The result of these effects necessitates much more complex analysis methods than what may be applicable in the single photon case, as capturing quantum nonlinear effects is essential. 

In this work, we utilize two complementary theory methods 
to study  two-photon correlations with temporal separations between the nontrivial photon wavepackets, in localized or de-localized form.
Specifically,
we make use of both a powerful tensor network approach using matrix product states (MPS)~\cite{PhysRevLett.116.093601}, which can be easily extended to larger numbers of input quanta for the Fock-state pulses, and a frequency-dependent scattering theory approach based on input-output theory. The scattering theory method has commonly been used to study CW interactions with a TLS, but has only seen limited use in the study of dynamics with pulsed quantized light~\cite{PhysRevLett.126.023603,FanMPSComparison}.

The study of induced correlations between  scattered photons from the two-photon pulse has previously been studied, predominantly in the case of {\it indistinguishable} photons. One such example is the {\it bird-like} bunching statistics in transmission~\cite{le_jeannic_dynamical_2022,matias_2025,Nysteen2015}, caused by two-photon dynamical interaction with a single quantum dot in a semiconductor waveguide. Recently, other works have also taken interest in the scattering of two-photon product states from quantum dots, using the quantum dot to mediate entanglement of the scattered field~\cite{Meguebel2025FreqEntangle}.

\begin{figure}
    \centering
    \includegraphics[width=\linewidth]{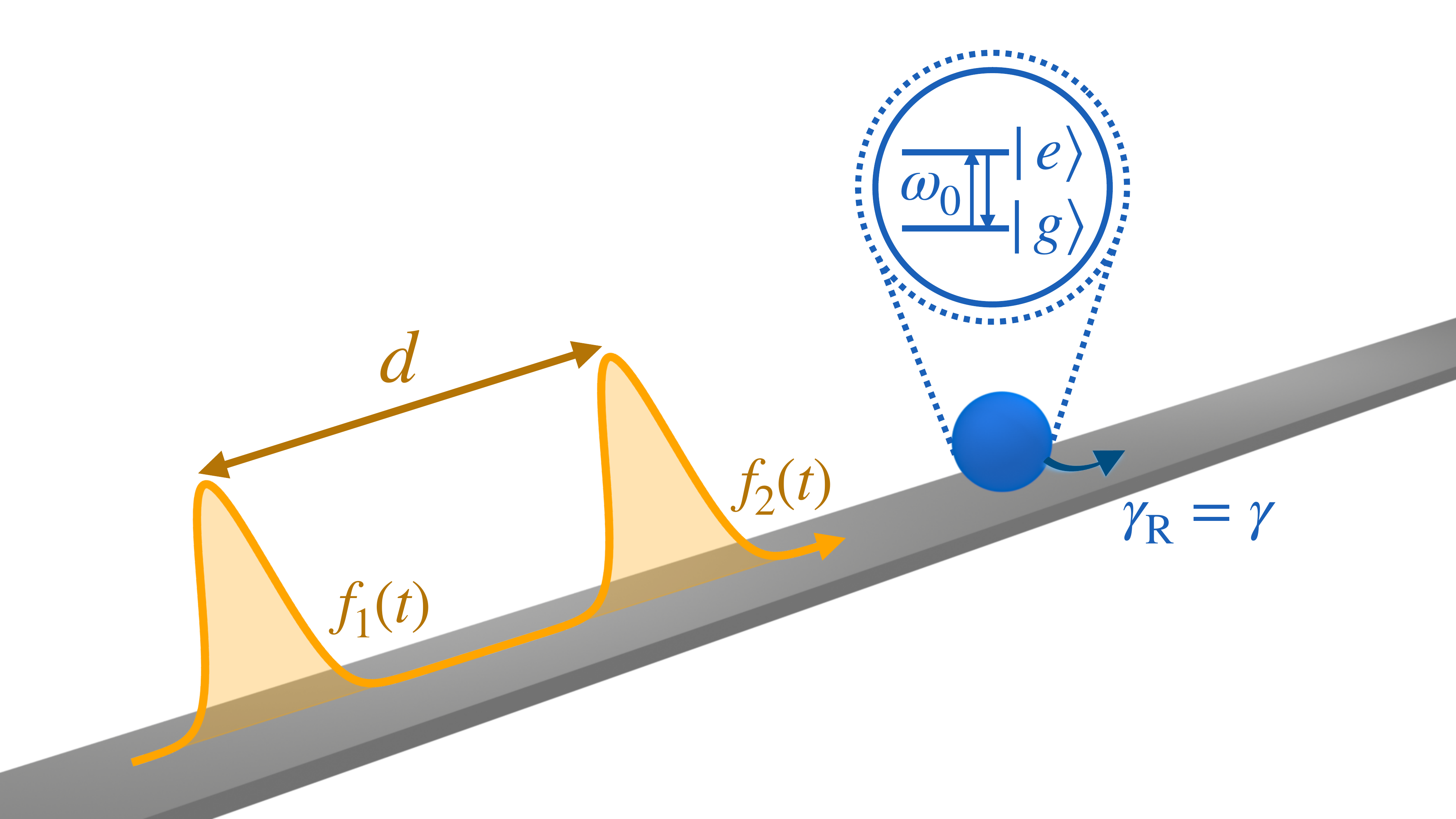}
    \caption{Schematic of the TLS chirally coupled (photon scattering is only in the forward/right direction) to the waveguide and an incident Fock pulse with a pulse envelope comprised of two spatial peaks, specified by $f_1(t)$ and $f_2(t)$ respectively. The spatial separation distance between these peaks, $d$, is given by $d=v_g t_b$, where $v_g$ is the group velocity of the waveguide mode. 
    }
    \label{fig:schematic}
\end{figure}

Another interesting aspect of two-photon pairs, whether entangled or not, is the additional degree of freedom that can be exploited with {\it time delays} in the pulse envelope, which is relatively easy to implement experimentally. 
Motivated by this extra degree of control, in this work we explore (photon and TLS) populations and 
field quantum correlations resulting from the scattering of two-photon states with bimodal pulse envelopes off a TLS (as depicted in \cref{fig:schematic}). We focus on the effects of varying the temporal correlations of the input state by differing localization conditions of the photons within the two-photon pulse envelope. For ease of understanding, we consider a chirally coupled TLS, though all of the methods used can be applied to symmetric emitters as well,
which we show in the Supplement~\cite{supp}.
Some aspects of two-photon temporal delays were considered in
\cite{PhysRevA.90.063832}, who studied phase-space quasiprobability distribution functions for photons as well as populations and long time spectra, for localized photon pairs.
In this work, 
we consider both localized and delocalize 
two-photon pairs as input bimodal field pulses, and show significantly different responses,
especially for the second-order quantum correlation function. 

Specifically, we study the significant differences in the scattering statistics between a {\it fully delocalized two-photon state} (labeled as $\ket{2}$) in which the two incident photons are {\it indistinguishable} from one another and both distributed equally among the pulse envelope peaks, as well as an {\it localized, partially distinguishable} two-photon state (labeled as $\ket{1}\ket{1}$)---in which one photon is localized within each of the peaks in the incident pulse envelope.
When the two peaks of the pulses overlap exactly, both photons are localized to the same space, and these two input states are identical and contain indistinguishable photons. We use the same-time second-order correlation function of the field as a signature of exiting the single photon subspace and inducing nonlinear scattering from the TLS. By comparing the transmitted correlation functions, we compare the stark differences between the scattering statistics dependent on the localization of photons, noting the impact that the separation of the two pulse peaks has on these dynamics. Using the scattering theory approach, we also consider scattering results from an interpolation between the fully temporally localized and delocalized photons in the input state, observing how slight delocalization immediately induces nonlinear scattering in the first pulse and the effects of delocalization on population transfer into the TLS.
Our findings are widely applicable to different experimental configurations presently being used in the community, from photon systems to circuit QED.

The rest of our paper is organized as follows. First, we introduce the relevant theory for the system Hamiltonian in \cref{sec:theory}, followed by a brief overview of the theory behind the MPS and frequency-based scattering theory approaches used to generate results. We then present results in \cref{sec:results}, first using top-hat temporal pulse envelopes for ease of analysis, and then followed by similar results in the case of Gaussian temporal pulse envelopes for realism. We continue in \cref{sec:Discussion} with discussing the experimental prospects of possible implementations to generate the 2-photon states of interest from single photon sources, and some of the physical implications and applications of this work.
Finally, in \cref{sec:conclusions} we conclude, summarizing our results and stressing the experimental feasibility of many of our observations.

\section{Theory}
\label{sec:theory}
In this section, we introduce the relevant theory used to produce the simulation results discussed later. We begin by introducing the Hamiltonian that is modeled, as well as the relevant assumptions in \cref{subsec:ham}. We then describe the background theory used to define this problem in a time-discretized basis using MPS in \cref{subsec:MPS}. Finally, we introduce the complementary frequency-dependent scattering theory approach in \cref{subsec:fanTheory}.

\subsection{Hamiltonian}
\label{subsec:ham}
The Hamiltonian of the waveguide-QED system describes a single TLS coupled to an infinite waveguide (in units of $\hbar=1$):
\begin{align}
    H = H_{\rm TLS} + H_{\rm WG} + H_{\rm I},
\end{align}
which includes contributions from the TLS, the waveguide, and their interactions respectively. The TLS Hamiltonian is given by
\begin{align}
    H_{\rm TLS}=\omega_A\sigma^+ \sigma^-,
\end{align}
where $\omega_A$ is the resonant frequency of the TLS transition, $\sigma^+$ is the Pauli operator transitioning the TLS from the ground state to the excited state, \mbox{$\sigma^+\equiv \ket{e}\bra{g}$}, and $\sigma^-$ is its Hermitian conjugate.
The free-field Hamiltonian inside the waveguide is given by
\begin{align}
    H_{\rm WG} = \sum_{\mu=L,R} \int_{-\infty}^\infty d\omega \omega b_\mu^\dag(\omega) b_\mu(\omega),
\end{align}
where \mbox{$\mu\in\{L,R\}$} denotes the left and right propagating channels of the field inside the waveguide, and $b_\mu(\omega)$ is a field operator for the photons in the waveguide satisfying the usual bosonic commutation relation \mbox{$[b_\mu(\omega), b_{\mu'}^\dag(\omega')] = \delta_{\mu,\mu'}\delta(\omega - \omega')$}. For convenience the bounds of the integral have been extended beyond the bandwidth of interest, to infinitely above and below the resonant frequency, to employ the first Markov approximation \cite{gardiner_zoller_2010}.

Lastly, the interaction Hamiltonian within the rotating wave approximation is given by
\begin{align}
    H_{\rm I} = &\sum_{\mu=L,R}\int_{-\infty}^\infty d\omega \left(\kappa_\mu(\omega)\sigma^+b_\mu(\omega) + \hc\right),
\end{align}
where $\kappa_\mu(\omega)$ is the coupling strength between the TLS and the light mode of frequency $\omega$ propagating in direction $\mu$. By transforming to the interaction picture in a rotating frame at $\omega_0$, where in the case of the TLS transition being resonant with the waveguide (\mbox{$\omega_A\equiv\omega_0$}), we can simplify the Hamiltonian to 
\begin{align}
    H = &\sum_{\mu=L,R}\int_{-\infty}^\infty d\omega\kappa_\mu(\omega)\sigma^+b_\mu(\omega)e^{-i(\omega-\omega_0)t} + \hc\,.
\end{align}

Defining the TLS decay rate as \mbox{$\gamma_\mu(\omega) = 2\pi\kappa_\mu^2(\omega)$}, and approximating this decay rate to be constant, we can express the Hamiltonian as
\begin{align}
    H = &\sum_{\mu=L,R} \sqrt{\gamma_\mu(\omega_0)}\left(\sigma^+b_\mu(t) + \hc\right),
\end{align}
where $b_\mu(t)$ satisfy the time domain bosonic commutation relation \mbox{$[b_\mu(t), b_{\mu'}^\dag(t')] = \delta_{\mu,\mu'}\delta(t - t')$}. Since we will consider a chiral TLS, we simply have $\gamma = \gamma_R(\omega_0)$ and $\gamma_L=0$, so that
\begin{align}
    H = \sqrt{\gamma}
    \left (\sigma^+ b_R(t) + \sigma^- b_R^\dag(t)\right ).
\end{align}

\subsection{Matrix Product States}
\label{subsec:MPS}
Simulating many-body quantum systems in a numerically exact way quickly becomes intractable due to the exponential scaling of the Hilbert space size. To address this problem, we use an MPS method developed for waveguide QED \cite{PhysRevLett.116.093601,PhysRevResearch.3.023030,arranzRegidor2026Qwavemps}, where we express our system in a time-discrete basis. 

The MPS basis is defined using time-bin noise operators, \mbox{$\Delta B_\mu(t_k)$}, which are related to the waveguide field bosonic operators by
\begin{align}
    \Delta B_\mu(t_k) = \int_{t_k}^{t_{k+1}} dt' b_\mu(t'),
\end{align}
which satisfy the bosonic commutation relation \mbox{$[\Delta B_\mu(t_k), \Delta B_{\mu'}^\dag(t_{k'})] = \Delta t \delta_{k,k'}\delta_{\mu, \mu'}$}, 
where \mbox{$\Delta t$} is the time increment of the simulation such that \mbox{$t_{k+1}-t_k=\Delta t$}. The eigenstates of the number operator, \mbox{$\Delta B_\mu^\dag(t_k) \Delta B_\mu(t_k)$}, define the basis at each time bin $k$. Using this formalism, the time evolution of the TLS and the incident time bin is given by
\begin{align}
    U(t_{k+1}, t_k) = \exp\left( -i\sqrt{\gamma}\,(\sigma^+ \Delta B_R(t_k) + \hc) \right).
\end{align}

Within this time-discrete basis, we then use the time-bin noise operators to evaluate photonic observables at time points $t$, such as the time-dependent transmitted flux,
\begin{align}
    n_T(t) = \frac{\braket{\Delta B_R^\dag(t) \Delta B_R(t)}}{(\Delta t)^2},
\end{align}
and the transmitted first and second order two-time correlation functions:
\begin{align}
    G_{TT}^{(1)}(t,\tau) =& \frac{\braket{\Delta B_R^\dag(t)\Delta B_R(t+\tau)}}{(\Delta t)^2},\\
    G_{TT}^{(2)}(t,\tau) =& \frac{\braket{\Delta B_R^\dag(t)\Delta B_R^\dag(t+\tau)\Delta B_R(t+\tau)\Delta B_R(t)}}{(\Delta t)^4},
\end{align}
where the subscript `$TT$' refers to two {\it transmitted} photons (both scattered to the right) for the second order correlation function.
In the case of symmetric coupling, analogous expectation values can easily be defined and evaluated for the reflected photon field. 

We will also find it convenient to define the shorthand for the same-time ($\tau=0$) second-order correlation function as
\begin{align}
    n_{TT} \equiv G_{TT}^{(2)}(t,0) = \frac{\braket{(\Delta B_R^\dag (t))^2(\Delta B_R(t))^2}}{(\Delta t)^4}.
\end{align}

Using this time-discrete basis with the MPS ansatz, we can express our total system state as
\begin{align}
    \ket{\psi} = \sum_{i_1,\dots i_m} \prod_{k=1}^m \left(A_k^{(i_k)}\right) \ket{i_1,i_2,\dots,i_m},
\end{align}
where each $A_k^{(i_k)}$ for $1<k<m$ is a rank 2 tensor, and $A_1^{(i_1)}$ is a rank 1 tensor that exists in the dual space of $A_m^{(i_m)}$ and each product over all of the $A_k$ results in a scalar. 
In the case of the MPS factorization of the coupled TLS-waveguide system state, we let the right most tensor/tensorspace be the TLS tensor space. Following Refs.~\cite{PhysRevA.103.033704,Guimond_2017,sofia2025}, we can operate on the vacuum state with time-bin noise creation operators at multiple time points to construct Fock state pulses expressed within this MPS formalism themselves. Doing so, we have the two-photon product state
\begin{align}
    \ket{\psi_{\rm in}} =& \frac{1}{\sqrt{2}}\iint dt_1 dt_2 f^{(1)}(t_1) f^{(2)}(t_2) b_R^\dag(t_1)b_R^\dag(t_2)\ket{0} \nonumber\\
    =& \frac{1}{\sqrt{2}}\sum_{i=1}^m \sum_{j=1}^m f^{(1)}(t_i) f^{(2)}(t_j) \Delta B_R^\dag(t_i) \Delta B_R^\dag(t_j)\ket{0} \nonumber\\
    =& \frac{1}{\sqrt{2}}\left(\sqrt{2}\sum_{i=1}^m f_i^{(1)}f_i^{(2)}\ket{2_i} + \sum_{i=1}^m \sum_{j=1}^m f_i^{(1)} f_j^{(2)}\ket{1_i,1_j}\right), \label{eq:twoPhotonState}
\end{align}
where we notate only the non-vacuum state sites in the final line of \cref{eq:twoPhotonState}. Here $f^{(j)}(t)$ is a time-dependent input pulse envelope that satisfies the normalization condition \mbox{$\int_{-\infty}^\infty dt |f^{(j)}(t)|^2 = 1$}, and $f^{(j)}_i$ is the discretization of this envelope, defined to be unitless and satisfying the analogous normalization condition \mbox{$\sum_{k=1}^N |f^{(j)}_k|^2 = 1$} that is ensured by having \mbox{$f^{(j)}_k\equiv f^{(j)}(t_k)\sqrt{\Delta t}$}. 

In this form, the quantum state can easily be expressed as an MPS factorization using language theoretic arguments \cite{bacon2008wfa}. Doing so yields the matrices for the 2-photon Fock state discretized over $m$ time bins, each of which has a physical dimension of 3 due to the maximum of two photons in a single bin, and a bond dimension of 4 to track the two distinct photon envelopes. These are explicitly given by (prior to transformation into a canonical form \cite{schollwock_density-matrix_2011}),
\begin{gather}
    A_1^{(0)} = \begin{pmatrix} 1 & 0 & 0 & 0\end{pmatrix}
    ,\qquad A_1^{(1)} = \begin{pmatrix} 0 & f_1^{(1)} & f_1^{(2)} & 0\end{pmatrix}
    ,\nonumber\\A_1^{(2)} = \sqrt{2} f_1^{(1)}f_1^{(2)} \begin{pmatrix} 0 & 0 & 0 & 1\end{pmatrix}, \label{eq:mps_fockMatrices_1}
\end{gather}
\begin{gather}
    A_k^{(0)} = \begin{pmatrix}
        1 & 0 & 0 & 0 \\ 0 & 1 & 0 & 0 \\ 0 & 0 & 1 & 0 \\ 0 & 0 & 0 & 1
    \end{pmatrix}
    ,\qquad A_k^{(1)} = \begin{pmatrix}
        0 & f^{(1)}_k & f^{(2)}_k & 0 \\
        0 & 0 & 0 & f^{(2)}_k \\
        0 & 0 & 0 & f^{(1)}_k \\
        0 & 0 & 0 & 0
    \end{pmatrix}
    ,\nonumber\\A_k^{(2)} = \sqrt{2}f_k^{(1)}f_k^{(2)}\begin{pmatrix}
        0 & 0 & 0 & 1 \\
        0 & 0 & 0 & 0 \\
        0 & 0 & 0 & 0 \\
        0 & 0 & 0 & 0 
    \end{pmatrix}, \label{eq:mps_fockMatrices_k}
\end{gather}
\begin{align}
    A_m^{(0)} = \begin{pmatrix} 0 \\ 0 \\ 0 \\ 1\end{pmatrix}
    ,A_m^{(1)} = \begin{pmatrix} 0 \\ f_m^{(2)} \\ f_m^{(1)} \\ 0 \end{pmatrix}
    ,A_m^{(2)} = \sqrt{2}f_m^{(1)}f_m^{(2)}\begin{pmatrix} 1 \\ 0 \\ 0 \\ 0 \end{pmatrix}, \label{eq:mps_fockMatrices_m}
\end{align}
where we have $1<k<m$. 

Similar matrices in a smaller subspace can be constructed to model single photon states, and prior to use, these matrices must be transformed to a canonical MPS form with the orthogonality center (OC) in the time-bin adjacent to the TLS.

With these representations of Fock states in the time domain, we can compose two kinds of input pulses of interest:
(i) two different temporal pulse envelopes sot that we have {\it two independent single photon pulses} with a photon localized within each pulse, denoted as the state $\ket{\psi}_{\rm in} = \ket{1}\ket{1}$; and 
(ii) an overall {\it two photon indistinguishable state} with both photons distributed over the pulse envelope given by the chain of time bins presented above and denoted as $\ket{\psi}_{\rm in} = \ket{2}$. 

In what follows, we will first use MPS to simulate two {\em top-hat} temporal pulse envelopes of identical amplitude, identical duration $t_p$, and with the pulse centers separated by some time $t_b$. This choice of pulse envelope is not a model restriction, but provides a clear starting point for analyzing the physics by presenting clear endpoints of the pulse effects and the beginning of purely spontaneous emission. It is also easier to implement in the MPS time domain than using pulsed scattering theory, as discussed below.
In the case of the two-photon state with $t_b \geq t_p$, this coincides with the input temporal pulse function
\begin{align}
    f_{\ket{2}}^{\rm th}(t) = \begin{cases}
        \frac{1}{\sqrt{2 t_p}}, & 0\leq t\leq t_p \text{ or } t_b \leq t \leq  t_b + t_p \\
        0, & \text{otherwise \ } 
    \end{cases}
\end{align}

In addition, for the two one-photon pulses, we have two subsequent pulses of pulse length $t_p$ given by
\begin{align}
    f_{\ket{1}\ket{1}}^{\rm th}(t) = \begin{cases}
        \frac{1}{\sqrt{t_p}}, & 0\leq t\leq t_p \\
        0, & \text{otherwise \ } 
    \end{cases},
\end{align}
of which we take $f^{(1)}(t) =  f_{\ket{1}\ket{1}}^{\rm th}(t)$ and $f^{(2)}(t) =  f_{\ket{1}\ket{1}}^{\rm th}(t-t_b)$. In both cases, we can see from \cref{fig:schematic} that the physical distance between the pulse centers is equal to the temporal distance, $t_b$, scaled by a factor of the propagating mode's group velocity.

Similarly, we can also define the two-photon temporal pulse envelope of two Gaussians separated temporally by $t_b$ as the sum of two Gaussians with means at times $t_c$ and $t_c+t_b$, and standard deviations in units of time, $\sigma_t$, given by
\begin{align}
    f_{\ket{2}}^{\rm g}(t) = C\left(e^{-\frac{(t-t_c)^2}{2\sigma_t^2}} + e^{-\frac{(t-(t_c+t_b))^2}{2\sigma_t^2}} \right), 
\end{align}
where $C$ is a normalization constant, given by
\begin{align}
    C^{-1} = {\sqrt{2\sigma_t\sqrt{\pi} \left( 1 + e^{ -\frac{t_b^2}{4\sigma_t^2}} \right)}}.
\end{align}

Once again, the temporal pulse envelopes used by two consecutive single photon pulses are two {\it independent}, square-normalized Gaussian pulse envelopes with respective means (pulse centers) $t_c$ and $t_c + t_b$, of the form
\begin{align}
    f_{\ket{1}\ket{1}}^{\rm g}(t) = \frac{1}{\sqrt{\sigma_t\sqrt{\pi}}}e^{-\frac{(t-t_c)^2}{2\sigma_t^2}}.
\end{align}

It should be noted that due to the simulation method used to define the temporal pulses in MPS, the input wave functions are restricted to two-photon product states. This is not a strict limitation of MPS, only one of the implementation, and by expanding the bond indices of the MPS factorization of the state, one may consider more complicated two-photon input states. Within the two-photon subspace of input states, the bond dimension, $\chi$, of the MPS remains tractable for arbitrary two-photon states, scaling as $2+2\chi_{\rm max}$ where $\chi_{\rm max}$ is the Schmidt rank of the input state's temporal wavefunction (or the necessary cut off for numerical precision). In the case of a Schmidt rank of 1, a product state, the bond dimension only has to be extended from 3 to 4 when compared to indistinguishable two-photon states. The scaling of the bond dimension is discussed in more detail in Sec.~S2 of the Supplement~\cite{supp}. In spite of this, it is technically easier to implement arbitrary input states using the frequency-dependent scattering theory method, as the state does not need to be factorized into an MPS.



\subsection{Frequency-Dependent Scattering Theory}
\label{subsec:fanTheory}

In this subsection, we briefly discuss a frequency-dependent scattering theory, in which the two-photon scattering matrix is used to solve for the asymptotic output fields exiting an initially ground state TLS \cite{PhysRevA.82.063821,xuInputoutputFormalismFewphoton2015a,PhysRevLett.126.023603}. A more complete derivation of the relevant theory involved in pulsed single and two-photon scattering is given in \cite{FanMPSComparison}. It can be shown that by Fourier transforming operators and observables, we can use the input/output operators in the frequency domain to describe time-dependent observables with this method---which is commonly applied to continuous wave systems \cite{PhysRevLett.126.023603}---to solve for pulsed solutions. The scattering theory approach makes use of input-output theory \cite{PhysRevA.31.3761}, where the input photon field is described using input bosonic operators which satisfy the commutation relation \mbox{$[a_{\rm in}(t) a_{\rm in}^\dag (t')] = \delta(t-t')$}. 

The output creation/annihilation operators describe the photonic field scattered by the (chiral) TLS and are related to the input bosonic operators by
\begin{align}
    a_{\rm out}(t) = a_{\rm in}(t) - i\sqrt{\gamma}\sigma^-(t).
\end{align}
These input-output operators can also be Fourier transformed to the frequency domain to yield frequency-dependent field operators of the form
\begin{align}
    a_{\rm in/out}(\omega) = \frac{1}{\sqrt{2\pi}} \int dt e^{i\omega t} a_{\rm in/out}(t).
\end{align}

Using the appropriate equations of motion for the system, one can determine the two-photon scattering matrix for the chiral system, which is derived to be
\begin{align}
    S_{\nu_1\nu_2\omega_1\omega_2} =& \bra{0}a_{\rm out} (\nu_1) a_{\rm out}(\nu_2) a_{\rm in}^\dag (\omega_1) a_{\rm in}^{\dag} (\omega_2)\ket{0} \nonumber\\
    =& t(\nu_1)t(\nu_2) [\delta(\nu_1 - \omega_1) \delta(\nu_2-\omega_2)\nonumber\\
    &+ \delta(\nu_1-\omega_2) \delta(\nu_2-\omega_1) ]\nonumber\\
    &+ T_{\nu_1\nu_2\omega_1\omega_2} \delta( \nu_1 + \nu_2 - \omega_1-\omega_2),
\end{align}
where \mbox{$t(\omega)$} is the single-photon transmission coefficient, which in the resonant chiral case is given by \mbox{$t(\omega) = \frac{-i\gamma/2 + \omega}{i\gamma/2 +\omega}$}, and 
\begin{equation}
    T_{\nu_1\nu_2\omega_1\omega_2} =  \frac{i\sqrt{\gamma}}{\pi} s(\nu_1) s(\nu_2) \left[  s(\omega_1) + s(\omega_2) \right],
\end{equation}
in which \mbox{$s(\omega) = \frac{\sqrt{\gamma}}{(\omega + i\gamma/2)}$} describes the nonlinear scattering contribution~\cite{PhysRevA.82.063821}. Without this final contribution, we have the two-photon scattering matrix for the quantum harmonic oscillator.

Using this frequency approach, the scattering theory can calculate field observables, such as two-time correlation functions and the transmitted flux of the output field described earlier, from a two-photon input state scattering off the two-level system. The photonic input state is specified in frequency space in the form
\begin{align}
    \ket{\psi_{\rm in}} = \frac{1}{\sqrt{2}}\iint d\omega_1 d\omega_2 f(\omega_1,\omega_2) a_{\rm in}^\dag(\omega_1)a_{\rm in}^\dag(\omega_2)\ket{0},
\end{align}
where $f(\omega_1,\omega_2)$ is the two-photon input spectra and satisfies the normalization condition \mbox{$\iint_{-\infty}^\infty d\omega_1 d\omega_2 |f(\omega_1,\omega_2)|^2 = 1$}. 

Conveniently, the frequency-dependent scattering theory provides direct access to the output scattered joint spectral amplitude
\begin{align}
    \ket{\psi_{\rm out}} = \frac{1}{\sqrt{2}}\iint d\omega_1 d\omega_2 f_{\rm out}(\omega_1,\omega_2) a_{\rm out}^\dag(\omega_1) a_{\rm in}^\dag(\omega_2)\ket{0},
\end{align}
allowing for direct analysis of the output state and comparison of it to the case of equivalent scattering off of a linear scattering emitter.

Since $f_{\rm out}(\omega_1,\omega_2)$ is symmetric, we can conduct a Schmidt decomposition to obtain the Autonne-Takagi factorization of the scattered state~\cite{houde2024}
\begin{align}
    f_{\rm out}(\omega_1,\omega_2) = \sum_k \sqrt{\lambda_k}u_k(\omega_1)u_k(\omega_2),
\end{align}
where $u_k(\omega)$ are the Schmidt modes, 
and we enforce the normalization of the Schmidt eigenvalues so that $\sum_k \lambda_k = 1$, and order the modes so that $\lambda_k \geq \lambda_{k+1}$. We can then use this factorization to calculate relevant changes to the scattered spectral space~\cite{law2000,miatto2012,ansari2018}, such as the Schmidt number
\begin{align}
    K = \frac{1}{\sum_k \lambda_k^2}, \label{eq:shmidtNum}
\end{align}
which is an inverse participation ratio and describes the number of Schmidt modes that contribute to the state. In particular, it is interesting to consider the change in the number of scattered contributing Schmidt modes resulting from the nonlinear scattering of the two-photon state off of a TLS when compared to a linear scattering process for the equivalent photonic input state, which we define as
\begin{align}
    \Delta K = K_{\rm nlin} - K_{\rm lin}, \label{eq:shmidtNumDiff}
\end{align}
where $K_{\rm nlin}$ is the Schmidt number of the scattered joint spectral amplitude resulting from the nonlinear scattering off of the TLS and $K_{\rm lin}$ is the Schmidt number of the scattered joint spectral amplitude resulting from linear scattering (equivalently the Schmidt number of the input two-photon state).

Additionally, we can also consider the fraction of the scattered state resulting from the nonlinear scattering process that exists within the space produced from the linear scattering process. To do so we determine the Autonne-Takagi factorization for the linear scattering process, and use this basis to construct the projector onto this linear scattered space,
\begin{align}
    P_{\rm lin} = \sum_{k=1} \ket{u_k}\bra{u_k},
\end{align}
so that it has components $(P_{\rm lin})_{ij} = \sum_ku_k(\omega_i)u_k^*(\omega_j)$. By projecting onto this space we can determine the fraction of the nonlinear scattered spectral amplitude that exists within the scattered linear subspace,
\begin{align}
    W_{\rm lin} = \frac{|| Pf_{\rm out}P^T ||^2}{||f_{\rm out}||^2},
\end{align}
where $||\cdot||^2$ denotes the Frobenius norm.

To compare analogous results to the input photon states from MPS, with either each photon completely temporally localized to a separate envelope or both photons distributed over the entire pulse envelope, we consider a separable input spectrum of the form :
\begin{equation}
f(\omega_1,\omega_2) = C[f_1(\omega_1)f_2(\omega_2) + f_1(\omega_2)f_2(\omega_1)],
\end{equation}
where $C$ is a normalization constant, $f_1$ and $f_2$ are the input single photon spectra, and $f$ is defined to be symmetric with respect to particle exchange of the photons. In the case of two Gaussians in the time domain, 
with standard deviation $\sigma_t$, and their means separated in the time domain by $t_b$, we then have the single-photon spectral functions 
\begin{align}
    g_1(\omega) =& e^{-\sigma_t^2 \omega^2/2 + i\omega t_c} \\
    g_2(\omega) =& e^{-\sigma_t^2 \omega^2/2 + i\omega(t_c + t_b)}.
\end{align}

To simulate the fully temporally-localized photons discussed previously with MPS as the input state, $\ket{1}\ket{1}$, where a single photon is localized to each Gaussian, we let $f_i(\omega) = g_i(\omega)$. To simulate the $\ket{2}$ input state, where both photons are identically distributed over the entire envelope, we let \mbox{$f_1(\omega)=f_2(\omega) = g_1(\omega) + g_2(\omega)$}. From the equations above, we can see that if $t_b=0$, these are the same initial states (indistinguishable), as the temporal localization of the photons is the only distinguishing feature between these two input states.

\begin{figure*}[h]
    \centering
    \begin{minipage}[t]{0.48\textwidth}
        \vspace{0pt}
        \centering
        \includegraphics[scale=0.95]{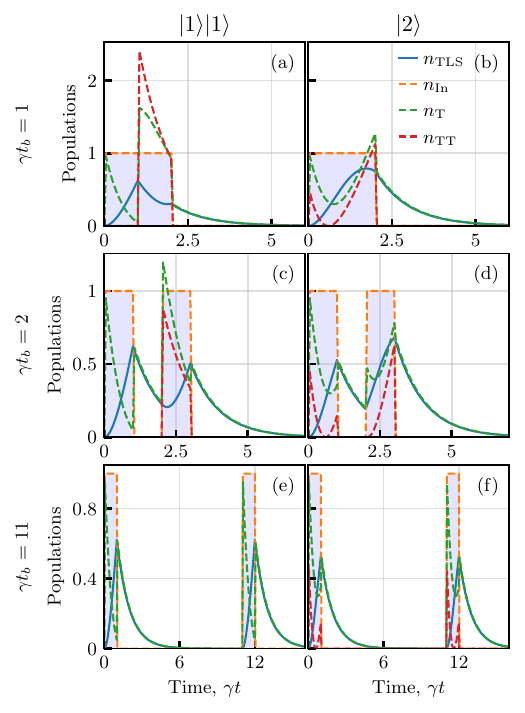}
        \caption{Population dynamics for two subsequent single photon top-hat pulses of length $\gamma t_p=1$ incident a chiral TLS-waveguide system with break times between the pulse centers of $\gamma^{-1}$ (a), $2\gamma^{-1}$ (c), and $11\gamma^{-1}$ (e) simulated with MPS. Results on the right side (b,d,f) are the same but for a single two photon pulse with the same temporal pulse envelope separated into two segments of length $\gamma^{-1}$.
        The blue shaded regions are to emphasize the pulsed area of the graph and the observables are defined by $n_{\rm TLS}=\braket{\sigma^+\sigma^-}(t)$, $n_{\rm in}$ being the input pulse flux, $n_T=\braket{\Delta B_R^\dag(t) \Delta B_R(t)}/(\Delta t)^2$, and $n_{TT} = G_{TT}^{(2)}(t,0)$.
        }
        \label{fig:mpsChiralTophatPops}
        \end{minipage}%
    \hfill
    \begin{minipage}[t]{0.48\textwidth}
        \vspace{6pt}
        \centering
        \includegraphics[scale=0.94]{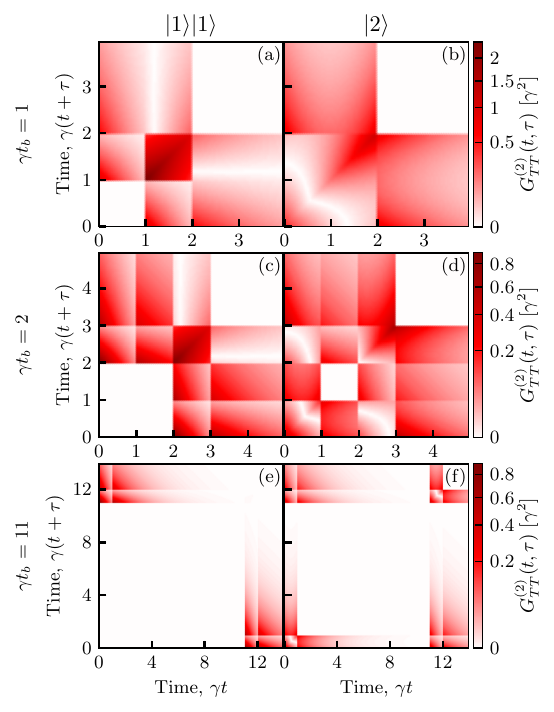}
        \caption{Two time point second-order correlation function in transmission, $G_{TT}^{(2)}(t,\tau)$ of the same simulations as \cref{fig:mpsChiralTophatPops} calculated with MPS. Plots are between two independent time points, $t$ and $t+\tau$.
        }
        \label{fig:mpsChiralTophatG2}
        \vfill
    \end{minipage}
\end{figure*}

Lastly, we note that the scattering theory formalism conveniently allows for complete control over the spectral profile of the input state (as a result of being a function of two frequencies). This allows a continuous analysis of the localization of the photons in the temporal envelope peaks, as we can define a {\it quantum-correlation mixing parameter} $\alpha\in[0,0.5]$, and take our single-photon spectral functions to be given by
\begin{align}
    f_1(\omega) =& (1-\alpha) g_1(\omega) + \alpha g_2(\omega) \label{eq:alpha_1}\\
    f_2(\omega) =& \alpha g_1(\omega) + (1-\alpha) g_2(\omega) \label{eq:alpha_2},
\end{align}
where $\alpha$ coherently tunes the localization of the single photon temporal modes via the weights of a superposition of the overall state.

It can easily be seen that for $\alpha=0$, we  then have the $\ket{1}\ket{1}$ state, as we have $f_i(\omega)=g_i(\omega)$ while for $\alpha=0.5$, we have equal contributions of both terms and have the $\ket{2}$ input state. Intermediate values of $\alpha$ then describe an interpolation between these conditions, with the photons being delocalized, but with probabilities favoring a single temporal Gaussian as well. 

Many of the previously described quantities are more difficult or impossible to calculate using MPS due to the output state being factorized as an MPS instead of having direct access to the output joint spectral amplitude. This demonstrates that there are subtle differences as well as advantages and disadvantages of using either MPS or scattering theory to explore two-photon scattering.


\begin{figure*}[t] 
    \centering\includegraphics[width=\textwidth]{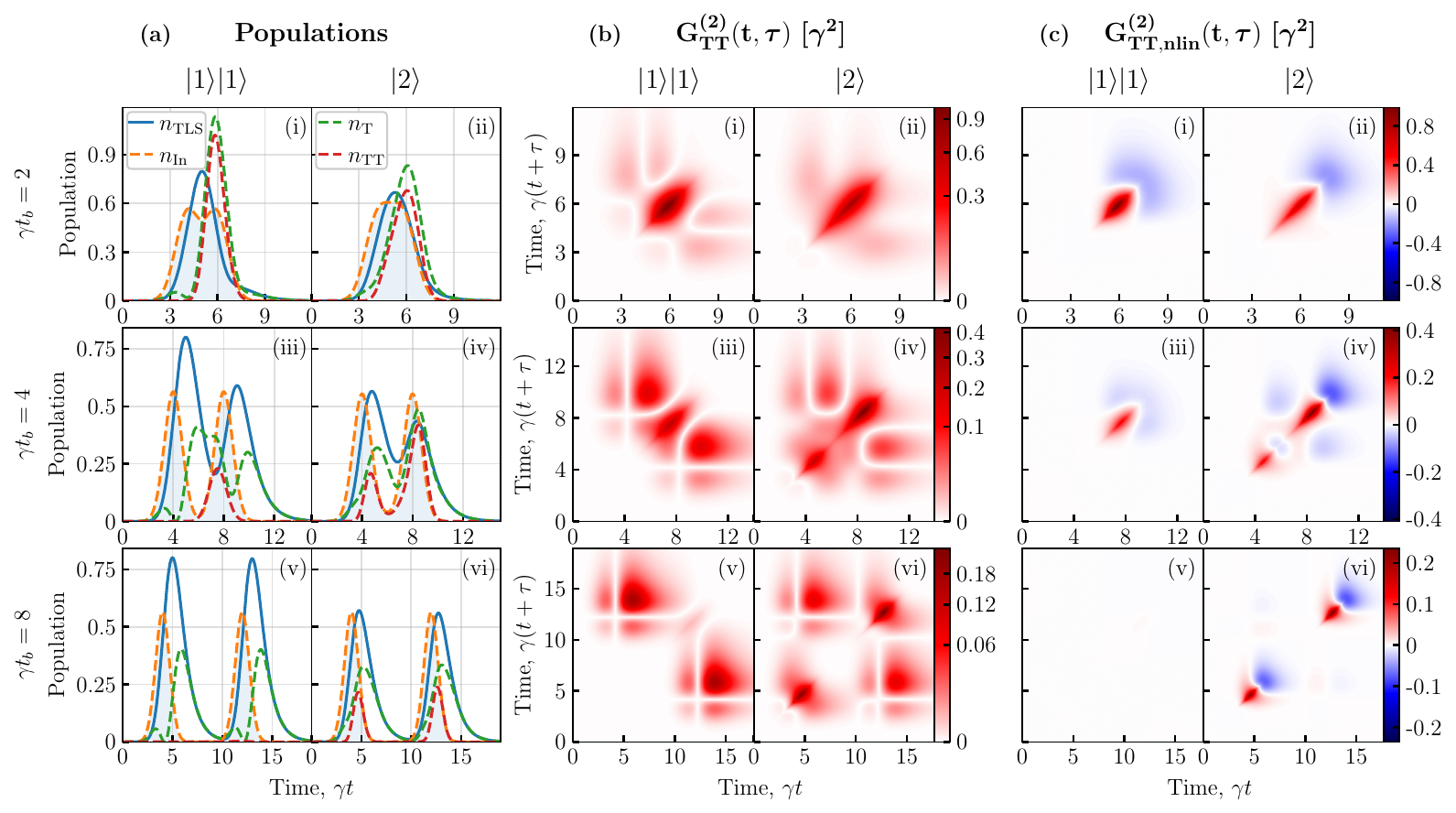}
    \caption{Dynamics of two subsequent single photon Gaussian pulses of standard deviation $\gamma\sigma_t=1$ and the first pulse centered at $\gamma t_c=4$ incident a chiral TLS-waveguide system with break times ($t_b$) between the pulse centers of $2\gamma^{-1}$ [a(i),b(i),c(i)], $4\gamma^{-1}$ [a(iii),b(iii),c(iii)], and $8\gamma^{-1}$ [a(v),b(v),c(v)] calculated using the scattering theory. Results on the right side for each group of graphs (ii,iv,vi) are the same but for a single two photon pulse into the TLS. Graphs in panel (a) show the population dynamics for with the blue shaded regions emphasizing the pulsed area of the graph and the observables are defined by $n_{\rm TLS}=\braket{\sigma^+\sigma^-}(t)$, $n_{\rm in}$ being the input pulse flux, $n_T=\braket{a_{\rm out}^\dag a_{\rm out}}(t)$, and $n_{TT} = G_{TT}^{(2)}(t,0)$. Panel (b) is the two time point second-order correlation function in transmission, $G_{TT}^{(2)}(t,\tau)$ of the same. Panel (c) is the nonlinear contribution to the second-order correlation function, $G_{TT,\rm{nlin}}^{(2)}(t,\tau)$.
    }
    \label{fig:fan_pop_G2}
\end{figure*}


\section{Results: Two-Photon Induced Dynamical Populations and Field Correlation Functions}
\label{sec:results}

In this section, we now showcase example population and first/second-order correlation function results produced using the previously discussed theory methods. First, for simplicity and clarity, we show the results from top-hat pulse envelopes in \cref{sec:topHats}, which are then followed by corresponding results with Gaussian pulse envelopes in \cref{sec:Gaussians}. In both cases, the population results were validated using analytical solutions from solving a hierarchical set of equation using input-output theory described in Sec.~S3 of the Supplement~\cite{supp}. We then conclude this section by using \cref{eq:alpha_1,eq:alpha_2} to compare results from two photon spectral functions that interpolate between the fully temporally localized and delocalized two photon states. 

\subsection{Temporal Top-hat Pulses}
\label{sec:topHats}


We observe that despite describing identical photonic fluxes into the TLS when using top-hat pulses (which is not the case for the Gaussians with overlapping envelopes), the $\ket{1}\ket{1}$ and $\ket{2}$ Fock pulses describe {\it radically different populations dynamics} [\cref{fig:mpsChiralTophatPops}]. Due to the temporal localization of the photons in the $\ket{1}\ket{1}$ case [\cref{fig:mpsChiralTophatPops}(a,c,e)], we see that for non-overlapping pulse envelopes, the initial pulse behaves linearly and with no $n_{TT}$ due to it exciting the TLS as a single photon \cite{sofia2025}. It is only upon the incidence of the second photon that nonlinear dynamics occur as the TLS can undergo 
{\it stimulated emission}; it is also at this point where $n_{TT}$ becomes nonzero as we exit the single quanta subspace. 

In contrast, the $\ket{2}$ input photon state [\cref{fig:mpsChiralTophatPops}(b,d,f)] has two quanta that are not localized within specific portions of the pulse envelope, resulting in {\it immediate nonlinear dynamics} of the TLS population and measurable same-time transmitted second-order photonic correlations. In both cases, we see that for sufficiently large separation times, the TLS returns to the ground state and the two pulse peaks behave as independent and identical excitation events; in the case of the $\ket{1}\ket{1}$ input state [\cref{fig:mpsChiralTophatPops}(e)], these are two independent single photon excitations of the TLS, and in the $\ket{2}$ input state case [\cref{fig:mpsChiralTophatPops}(f)], these are two identical and independent portions of the photonic pulse with a cosine squared modulation to the spectral properties arising from the temporal gap between the pulse peaks.

Corresponding to these single-time dynamics, we have the second-order correlation functions in transmission shown in \cref{fig:mpsChiralTophatG2}. In the case of the $\ket{1}\ket{1}$ input state [\cref{fig:mpsChiralTophatG2}(a,c,e)], this figure emphasizes the localization of the photons as $G_{TT}^{(2)}(t,\tau)=0$ for all $t,t+\tau<t_b$ due to the physics of the system behaving within a single quanta subspace prior to the incidence of the second photon. It is only upon the incidence of the second pulse, at time $t_b$, that the second-order correlation function becomes nonzero. For $t_b>t_p$, we can see from \cref{fig:mpsChiralTophatG2}(c) 
that four distinct time periods describe the dynamics, the initial pulse, the break between the pulses, the second pulse, and the spontaneous decay of the TLS after all pulse activity. For small $t_b$, we observe strong bunching during the second pulse as the TLS has not yet decayed upon the incidence of the second pulse, resulting in a high probability of the two photons being transmitted during this time period. For sufficiently large $t_b$, the two-time correlation function is simply that of single photon excitation and subsequent spontaneous emission of two independent TLS events.

Turning again to the two-photon correlated pulse, i.e., a $\ket{2}$ input state
[\cref{fig:mpsChiralTophatG2}(b,d,f)], since the photons are not temporally localized and are distributed over the entire pulse envelope, there is immediately nonzero $G_{TT}^{(2)}(t,\tau)$. It is only within the single photon subspace from the spontaneous emission of the TLS that we observe an abrupt shift to $G_{TT}^{(2)}(t,\tau)=0$, where $t_p < t,t+\tau < t_b$ and $t, t+\tau > t_p + t_b$. In the long $t_b$ limit [\cref{fig:mpsChiralTophatG2}(f)], 
the $G_{TT}^{(2)}$ still has the non-zero regions associated with spontaneous emission of the TLS during the two different pulses analogous to the $\ket{1}\ket{1}$ case, but with additional distinct two photon effects during each of the pulses due to possible presence of both photons in either segment of the pulse.

\subsection{Gaussian Pulses}
\label{sec:Gaussians}

\begin{figure}[t] 
    \centering
    \includegraphics[width=\columnwidth]{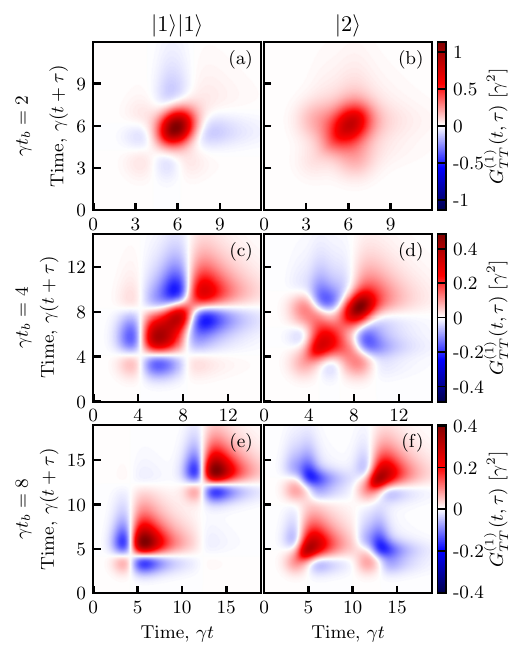}
    \caption{First order correlation function, $G_{TT}^{(1)}(t,\tau)$ for two subsequent single photon Gaussian pulses of standard deviation $\gamma\sigma_t=1$ and the first pulse centered at $\gamma t_c=4$ incident a chiral TLS-waveguide system with break times ($t_b$) between the pulse centers of 
    $2\gamma^{-1}$ (a), $4\gamma^{-1}$ (c), and $8\gamma^{-1}$ (e) calculated using the scattering theory. Results on the right side (b,d,f) are the same, but for a single two-photon pulse incident on the TLS.
    }
    \label{fig:fanChiralG1}
\end{figure}

\begin{figure*}[t] 
    \centering
    \includegraphics{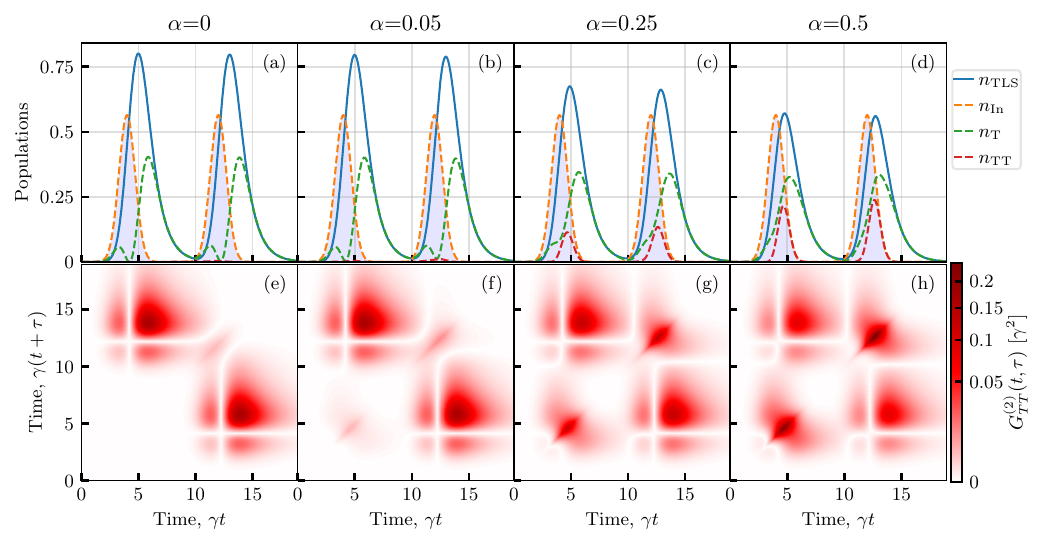}
    \caption{Population dynamics and second-order correlation function $G_{TT}^{(2)}(t,\tau)$ for pulse envelopes defined spectrally using two Gaussians and \cref{eq:alpha_1,eq:alpha_2} with standard deviations of $\gamma\sigma_t=1$, the first pulse center at $\gamma t_c=4$, and temporal distance between the peaks of $\gamma t_b = 8$. These results were calculated with the frequency-space scattering theory method and emphasize how the localization parameter, $\alpha$, interpolates between the $\ket{1}\ket{1}$ and $\ket{2}$ cases of input states.
    }
    \label{fig:fanChiralGaussiansAlpha}
\end{figure*}

The scattering theory approach was used to simulate the two-photon states with Gaussian pulse envelopes (correlations and populations were also validated with MPS). 

In this Gaussian pulses case, both for the population dynamics and for the second-order correlation function [see \cref{fig:fan_pop_G2}(a-c)], we observe analogous results to the top-hat pulse envelopes but without the jump discontinuity associated with the discrete change in those pulse envelopes. Here, we  emphasize the distinct effect of the delocalization of the two photons, such as a tighter and higher transmitted flux distribution in the short and long $t_b$ regimes [\cref{fig:fan_pop_G2}a(i,v)], and a consistently larger maximum excitation of the TLS than the $\ket{2}$ case. We can also see that the greatest complexity in the dynamics occurs in the intermediate regime in which the Gaussians have only a slight temporal overlap [\cref{fig:fan_pop_G2}a(iii,iv)], as strongly overlapping pulses might be approximated by an input pulse with identical photons [\cref{fig:fan_pop_G2}a(i,ii)], and pulses with large separations behave as two independent excitation events with the TLS starting in the ground state [\cref{fig:fan_pop_G2}a(v,vi)]. 

It is also worth noting that with the temporal overlap of the pulses (which was not present in the top-hat case), for small separations of the pulses [\cref{fig:fan_pop_G2}a(i,ii)], we can observe distinct and slight differences in the input flux between the $\ket{1}\ket{1}$ and $\ket{2}$ states. This is due to the $\ket{2}$ state containing indistinguishable photons, including their localizations, so that the flux is trivially $n_{\rm in} {\propto} |f_1(\omega)|^2 = |f_2(\omega)|^2$. In contrast, the differences in the photons in the $\ket{1}\ket{1}$ case cause unequal contributions from the combinations of products of the single-photon envelopes.

Regarding the second-order correlation functions for the $\ket{2}$ input state, in the case of a short separation time between the pulse peaks [\cref{fig:fan_pop_G2}b(ii)], we initially only observe faint signatures of the familiar `bird-like' correlation, as the significant (but not complete) overlap of the two Gaussians cause their sum to vary significantly from that of a single Gaussian. In the intermediate overlap regime, this novel correlation becomes clearer, particularly in the bottom left of the graph, though there is still strong interference effects between the two pulse peaks. 

Finally, when there is sufficiently large time between the two pulse peak incidences [\cref{fig:fan_pop_G2}b(vi)], we can clearly observe along the graph's diagonal two instances of the `bird-like' correlation pattern observed experimentally~\cite{le_jeannic_dynamical_2022} as well as in other analyses of 2 photon scattering without temporal delays~\cite{matias_2025,Nysteen2015,FanMPSComparison}. 
The dynamics of these correlations are analogous to those seen with top-hat pulses, where the differences between the two cases are most strongly emphasized with a slight temporal overlap between the two pulses [\cref{fig:fan_pop_G2}b(iii,iv)]. As the pulses get close to a complete overlap, the dynamics due to the $\ket{1}\ket{1}$ and $\ket{2}$ states become more similar [\cref{fig:fan_pop_G2}b(i,ii)]. 
For large separation distances of the pulse peaks, we observe the similar single photon off-diagonal time effects again due to linear scattering, while the $\ket{2}$ has strong correlations about the same time due to the probability of both pulse peaks containing two photons with the weaker off diagonal two photon probability due predominantly to linear scattering [\cref{fig:fan_pop_G2}b(v,vi)]. 

These delay-modified temporal correlation
effects are further emphasized in \cref{fig:fan_pop_G2}c(i-vi), which isolates the nonlinear contribution to the second-order correlation function by subtracting off the result due to purely linear scattering. This demonstrates the dominance of the nonlinear contribution along the diagonal of equal time points. Moreover, it makes clear the significance of the separation distance on the nonlinear interaction and the gradual shift to purely linear interaction in the case of the localized photon input states \cref{fig:fan_pop_G2}c(i,iii,v). Additionally, if we integrate over the entire two dimensional map of the absolute value of the nonlinear contribution to the second-order correlation, $\iint dt d\tau |G^{(2)}_{\rm TT,nlin}(t,\tau)|$, we observe both a consistent reduction in the magnitude of nonlinear correlation changes for larger $t_b$, and more extremes in the $\ket{1}\ket{1}$ case, as \cref{fig:fan_pop_G2}c(i) has a greater magnitude of nonlinear correlations than \cref{fig:fan_pop_G2}c(ii).

We highlight that the strong zero lines present on the off-diagonals of \cref{fig:fan_pop_G2}b, where linear contributions of the scattering processes dominate, are due to the presence of a root in the convolution of the single photon temporal pulse envelope with the Fourier transform of the chiral transmission coefficient, $\mathcal{F}[t(\omega)]$. In the case where $\gamma\sigma_t=1$, these roots appear slightly after $\gamma t_c$ and $\gamma(t_c+t_b)$, though this is not the case for wider pulses. These zero lines disappear in some of the graphs as they approach the diagonal, where nonlinear scattering contributions may be significant. This dependence on the pulse envelope is why such zero lines are not so prominent in the case of the top-hat pulses seen in \cref{fig:mpsChiralTophatG2}.


A similar explanation with neglecting nonlinear contributions can be used to explain the sharp, straight zero lines observed in the first-order correlation functions for the localized photons in the $\ket{1}\ket{1}$ state, particularly for large $t_b$, as seen in \cref{fig:fanChiralG1}. In fact, when the two photons are localized to pulse envelopes 
with sufficiently large separation time between the pulses, the scattering process is strictly linear due to the excitation of the TLS being restricted within the single photon subspace. 

This can be recognized by comparing \cref{fig:fan_pop_G2}b(v) and \cref{fig:fanChiralG1}(e). In this limit, the second-order correlation function is in fact equivalent to the time translated square of the first-order correlation function, where the same-time features of the first-order correlation must have one time coordinate translated by $t_b$ to coincide with the off-diagonal features of the second-order correlation function. 
The other cases are nontrivial due to significant contributions from nonlinear scattering processes. These are most notable in the first-order correlations in how they modify the features resulting from the $\ket{2}$ input state, such as curving the zero lines due to the interference between the linear and nonlinear contributions. 

While thus far we have considered only fully localized or delocalized photons, we note that we can use \cref{eq:alpha_1,eq:alpha_2} to interpolate between the results of these two kinds of pulses via a superposition, as demonstrated in \cref{fig:fanChiralGaussiansAlpha} using the temporal Gaussian pulse envelopes 
from previous results. For $\alpha=0$ and $\alpha=0.5$ [\cref{fig:fanChiralGaussiansAlpha}(a,d,e,h)], we can see that the results are exactly the same as those demonstrated above for the $\ket{1}\ket{1}$ and $\ket{2}$ input states, respectively, as these states coincide with the single photon temporal modes being fully localized/delocalized in the pulse envelopes respectively. In the temporal populations, we can see that as $\alpha$ increases, the transmitted flux decreases in height and broadens as the distributions converge upon the results of the $\ket{2}$ state. Likewise, the delocalization resulting immediately from $\alpha\neq0$ is evident in \cref{fig:fanChiralGaussiansAlpha}(f) as even with only $\alpha=0.05$ there is non-zero $G_{TT}^{(2)}(t,\tau)$ signatures with both time points during the first pulse, signifying an exit from the single photon subspace. This is because the $\alpha$ mediated 2-photon input state can be expressed as a superposition of the jointly localized and delocalized single photon modes, so as $\alpha$ increases from 0 to 0.5 the fraction of the state described by two indistinguishable delocalized single photon modes increases, which allows for probability of 2-photon scattering.

\begin{figure}[t] 
    \centering\includegraphics[trim={0.1cm 0.cm 0.1cm 0},clip,width=\columnwidth]{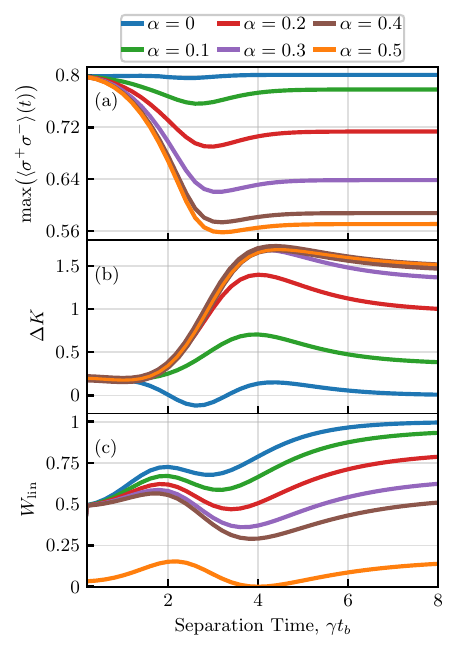}
    \caption{(a) Maximum TLS population, $\braket{\sigma^+\sigma^-}(t)$,
    (b) change in the Schmidt number, $\Delta K$, and (c) the weight within the linear scattering subspace, as a function of the temporal distance between peaks, $\gamma t_b$, for pulse envelopes defined spectrally using two Gaussians and \cref{eq:alpha_1,eq:alpha_2} with standard deviations of $\gamma\sigma_t=1$, and various values of $\alpha$. These results were calculated with the frequency-space scattering theory method.}
    \label{fig:scat_modes_analysis}
\end{figure}

In the population trends shown in \cref{fig:fanChiralGaussiansAlpha}, we can also see that, as $\alpha$ increases, the population peaks gradually reduce in their maximum achieved population, as corroborated by the two columns of \cref{fig:fan_pop_G2}a. We can observe that this is a broader phenomenon that holds for all separation distances of the two pulse peaks $t_b$, given Gaussian pulses with $\gamma\sigma_t=1$, as demonstrated in \cref{fig:scat_modes_analysis}(a). Here, we see that increasing $\alpha$ causes a systematic reduction in the maximum TLS population, which is due to the two-photon content of the pulses inducing immediate stimulated emission,
with a detrimental impact on the transfer of population to the TLS. Additionally, we observe the expected result of the maxima reaching steady-state values as $t_b$ becomes sufficiently large such that the dynamics behave as two independent excitations of the TLS ($\gamma t_b \gg 1$). 

Interestingly, the flatness of the $\alpha=0$ curve indicates that a nearly identical maximum TLS population is reached, whether one considers the excitation of a single Gaussian pulse containing two indistinguishable photons or a Gaussian pulse containing a single photon (in the $\gamma\sigma_t=1$ case). However, it should be noted that, while the maximum TLS population does not change significantly for varying $t_b$ with $\alpha=0$, the population dynamics are radically different, as demonstrated in \cref{fig:fan_pop_G2}a(i,iii,v). 
We can also see that in the large $t_b$ limit, the $\alpha=0.5$ case induces a maximum TLS population equal to the $\gamma t_b=0$ case divided by $\sqrt{2}$, which follows from the two-photon pulse being split into two separate excitation events of equal strength. 

It is also noteworthy that the maximum TLS population reaches a global minimum for each curve with a separation distance between the pulse peaks between $\gamma t_b=2$ and $\gamma t_b=4$. This is the result of two competing influences on the TLS population presented in Sec.~S3 of the Supplement~\cite{supp}, the positive contribution due to the population from the pulse that excites the TLS, $P^{\rm(pulse)}(t)$, and a negative contribution due to saturation of the TLS, $P^{\rm(sat)}(t)$. Both contributions grow in concentration as $t_b$ decreases, as the incident pulses become more overlapped to pump the TLS, but also results in saturation effects that inhibits excitation by the second photon. Increasing the pulse separation slightly, on the order of the TLS decay/pulse width, results in a more significant reduction in the positive contribution to the population, while the inhibiting saturation effects persist due to the TLS remaining partially excited between the two pulse peaks. The result is that in this regime these two quantities balance to inhibit the population transfer to the TLS and create a minima in the max TLS population. As $t_b$ is further increased the TLS can return to its ground state between the pulse peaks, greatly reducing the influence of $P^{(sat)}(t)$ and allowing more population into the TLS. It is also worth noting that it is only for slightly larger $t_b$ that the maximum amount of the scattered state exists outside of the linear scattering manifold [\cref{fig:scat_modes_analysis}(c)], supporting the strength of saturation resulting in nonlinear scattering in this regime.

From \cref{fig:scat_modes_analysis}(b) we see the change in the Schmidt number resulting from the nonlinear scattering, which describes the relative change in the number of Schmidt modes with meaningful occupation. We observe that when the photons are localized to their individual pulse envelopes ($\alpha=0$),  despite there being significant nonlinear contributions to the correlations for small and mid values of $t_b$ [\cref{fig:fan_pop_G2}c(i,iii)], we observe that the number of occupied modes is left relatively unchanged, with even the possibility of marginal consolidation into fewer modes in the intermediate regime \cref{fig:scat_modes_analysis}(b). By referencing \cref{fig:scat_modes_analysis}(c), we can see that within this nonlinear scattering regime a modest portion of the scattered state exists outside of the linear scattering space, demonstrating a rotation of these dominant Schmidt modes rather than a change in the size of the space through any significant mode generation. In contrast, we see that when more significant portions of the state is delocalized (as $\alpha$ increases), there is stronger generation of modes and a greater a greater portion of the scattered state existing outside of the linear scattering space, even when the temporal correlations appear similar \cref{fig:fan_pop_G2}c(i,ii). As trivially expected, in the limit of large $t_b$ the localized photons case ($\alpha=0$) produces no additional modes and the scattered state exists entirely in the space of the linear scattering modes, as it is equivalent to linear scattering.

In \cref{fig:scat_modes_analysis}(c),  we note that, in the $\alpha=0.5$ case the input state (and as such the linearly scattered state), has a Schmidt rank of one, resulting in a significantly smaller fraction of the nonlinear scattered norm existing within this smaller dimensional space. Alternatively, if we wanted to project into a more similarly sized basis we could instead determine the weight of the norm existing in the space given by the projector in the $\alpha=0.4$ case; in this example, the curve would lie just below the $\alpha=0.4$ curve and follow the same general trend already seen in the graph, with a further reduction of the state's norm in the linear space for increasing $\alpha$.



\section{Discussion}
\label{sec:Discussion}
As mentioned earlier, these simulations 
use parameters that are currently experimentally accessible.
For instance, a typical decay rate of a quantum dot system is of the order of $\gamma=1 \ {\rm ns^{-1}}$~\cite{doi.org/10.1002/qute.201900021}. Thus, if we consider a delay between pulses of $\gamma t_b =2$, then $t_b \approx 2 \ {\rm ns}$. Recent experiments in this regime have used pulses of an order of magnitude shorter~\cite{Mnaymneh2019}. Furthermore, in the case of working with superconducting qubits, we can consider a $\gamma/2\pi=17 \ {\rm MHz}$~\cite{gd4s-fgwt}, and for $\gamma t_b =2$, we would have $t_b \approx 19 \ {\rm ns}$. Pulses on the same time scale are already being used in experiments~\cite{PhysRevX.4.041010}.

For the preparation of the first initial input state, denoted as $\ket{1}\ket{1}$, the required setup is two synchronized single photon sources on demand with the necessary pulse reshaping for the single photons emerging from their sources. In this way, each photon can be generated with the form
\begin{align}
    \ket{1_{g_1}} = A_{\rm in}^\dag[g_1]\ket{0},
\end{align}
where
\begin{align}
    A_{\rm in}^\dag[g_i] = \int dt g_i(t)a^\dag(t),
\end{align}
is the single photon creation operator that creates a photon with the envelope $g_i(t)$. Since $g_2(t) = g_1(t-t_b)$, by adding a delay line to the channel of one of the sources prior to them joining the same channel that is incident the TLS we can introduce the inhomogeneous timing and the delay parameter $t_b$.

Similarly, the $\alpha$ parameter can be introduced via a stochastic state preparation technique by using a Mach-Zehnder interferometer with one arm lengthened to introduce the necessary delay line. Using such a device one could generate the two output channels

\begin{align}
    a_{\rm in,1}^\dag = \frac{1}{\sqrt{2}}\left( \sqrt{1-\eta}A_{\rm in}^\dag[g_1] +\sqrt{\eta}A_{\rm in}^\dag[g_2] \right) \\
    a_{\rm in,2}^\dag = -\frac{i}{\sqrt{2}}\left( \sqrt{1-\eta}A_{\rm in}^\dag[g_1] -\sqrt{\eta}A_{\rm in}^\dag[g_2] \right),
\end{align}
where we have introduced
\begin{align}
    \eta = \frac{\alpha^2}{(1-\alpha)^2 + \alpha^2},
\end{align}
so that $a_{\rm in,1}^\dag$ is the desired state. We stress that since $g_1$ and $g_2$ may not be orthogonal, and the normalization of the state is maintained by the appropriate normalization constant, $C$.

Alternatively, electro-optical techniques could be used to directly route~\cite{svarc_19} or reshape~\cite{kolchin_2008} the pulses and avoid stochastic production of the state resulting from neglecting the output from one of the arms of the Mach-Zehnder interferometer. Other techniques have also been developed for the general shaping of single photon temporal wave functions~\cite{morin_2019,karpinski2021}.

The same procedure could be conducted with the other single photon source, with the long and short arms swapped to generate the oppositely weighted contributions to $f_2$.

Lastly, we highlight the potential interest and importance of nonlinear scattering processes for quantum optical applications, such as for the implementation of gates for photonic quantum computing and as a nonlinear source in the development of quantum neural networks~\cite{heuck:2020,yanagimoto:2022,steinbrecher:2019,ewaniuk:2023,vazquez:2026}. Additionally, it is known that the presence of two-photons is important for these applications \cite{banacloche:2010,shapiro:2006}. This work  highlights the use of photon localization and delays as an additional degree of freedom that can be manipulated to modify the nonlinear interaction of a two-photon state with a qubit, and  also demonstrates the possible detrimental effects to the nonlinear interaction that accumulated timing delays between photons in cascaded systems might produce.

\vspace{0.2cm}
\section{Conclusions}
\label{sec:conclusions}

We have shown the significant influence of delay-induced temporal delocalization of a pulsed two-photon incident state on a TLS by computing the population dynamics and the second-order correlation function of the transmitted field using numerically exact approaches (MPS and scattering theory). We observed how two-photon Fock states with similar or identical incident fluxes may result in dramatically different dynamics, first using top-hat pulses as an illustrative example to describe how the time points of the incidence of the pulses impact the dynamics; then using Gaussian pulses as a more practical physical example.

We continued by noting a relationship between the first and second order quantum correlations in the fully linear scattering regime. We also highlighted the impacts of the nonlinear scattering contribution using the first and second order correlations and comparing to the near-linear scattering results generated by the $\ket{1}\ket{1}$ state with the pulse peaks widely separated. We also subtracted off the linear scattering contribution to the second order function to fully isolate the nonlinear scattering contribution to the scattered photon statistics.

Subsequently, we studied scattering from an interpolation between the two regimes of completely temporally localized and delocalized two photon states, noting that any amount of delocalization of the photons breaks the restriction of the dynamics being limited to the single photon subspace during the initial pulse and resulting in initial nonzero second order correlation signatures. We then considered how the combination of variables of the photon localization and pulse center separations modify the maximum excitation achieved by the TLS, noting that decreasing photon localization systematically reduces the maximum excitation achieved by the TLS (in the case of Gaussian pulses with $\gamma\sigma_t=1$). Lastly we used a Schmidt decomposition of the scattered state to analyze the generation of Schmidt modes from the nonlienar scattering process, and how the nonlinear scattering changes the space of the scattered state.

Our results can easily be extended to symmetrically coupled TLSs, with both transmitted and reflected field observables (with select results demonstrated in Sec.~S1 of the Supplement~\cite{supp}), as well as multiple TLSs. Moreover, the MPS simulations could also easily be extended to higher incident number states or more complicated initial conditions of the TLS that are not accessible via the scattering theory approach, which is practically limited to the two excitation subspace (as presented in \cite{FanMPSComparison}). However, as implemented the MPS approach is still limited by its specification of the temporal pulse envelope being a product state, not of the complete $N$ photon spectra function, resulting in the specification of only input states with a Schmidt rank of one. This is not a strict limitation of the method, as by conducting a Schmidt decomposition of the input state and constructing the appropriately structured matrices with larger bond indices it is possible to simulate arbitrary two photon states (though in principal this is still limited computationally by memory by the entanglement of the state). Thus, general two photon states with entanglement are significantly more challenging to implement with MPS, and if the discretized two photon temporal pulse envelope has too high of a Schmidt rank it may be intractable. 

Lastly, we emphasize that many of the results produced here are experimentally accessible with current day technologies. Similar experiments have been performed measuring the second order correlation function of two photons scattering from a 
semiconductor quantum dot waveguide system, though these measurements were limited to delocalized two photon excitations or widely separated single photons (with $\gamma t_b\gg 1$) \cite{le_jeannic_dynamical_2022}. Related experiments can also be implemented in waveguide schemes with circuit QED, though directly measuring the multi-photon correlation functions is more challenging in the microwave regime.

The MPS results presented in this paper were simulated using the open source Python package QwaveMPS \cite{arranzRegidor2026Qwavemps}.

\begin{backmatter}

\bmsection{Acknowledgments}
This work was supported by the Natural Sciences and Engineering
Research Council of Canada (NSERC), 
the National
Research Council of Canada (NRC), the Canadian Foundation
for Innovation (CFI), and Queen’s University, Canada.

\bmsection{Disclosures}
The authors declare no conflicts of interest.

\bmsection{Data availability}
Data underlying the results presented in this paper are
not publicly available at this time but may be obtained
from the authors upon reasonable request.

 \bmsection{Supplemental document}
 See Supplement 1 \cite{supp} for supporting content. 
\end{backmatter}


\bibliography{references_all}
\end{document}


\maketitle

\section{Symmetrically Coupled Two Level System  Results}

\begin{figure}[h] 
    \centering
    \includegraphics[width=0.65\linewidth]{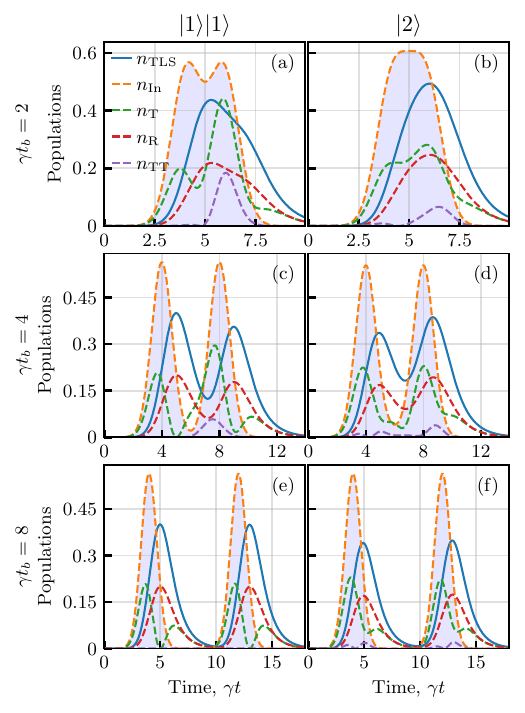}
        \caption{Population dynamics for two subsequent single photon Gaussian pulses of standard deviation $\gamma\sigma_t=1$ and the first pulse centered at $\gamma t_c=4$ incident a symmetric TLS-waveguide system with break times ($t_b$) between the pulse centers of $2\gamma^{-1}$ (a), $4\gamma^{-1}$ (c), and $8\gamma^{-1}$ (e) calculated using the scattering theory. Results on the right side (b,d,f) are the same but for a single two photon pulse into the TLS. The blue shaded regions are to emphasize the pulsed area of the graph and the observables are defined by $n_{\rm TLS}=\braket{\sigma^+\sigma^-}(t)$, $n_{\rm in}$ being the input pulse flux, $n_T=\braket{a_{\rm out,R}^\dag a_{\rm out,R}}(t)$ being the transmitted flux, $n_R=\braket{a_{\rm out,L}^\dag a_{\rm out,L}}(t)$ being the reflected flux, and $n_{TT} = G_{TT}^{(2)}(t,0)$.}
    \label{fig:fan_sym_pops}
\end{figure}

\begin{figure}[H]
    \centering
    \begin{minipage}{0.495\linewidth}
        \centering
        \includegraphics[width=\linewidth]{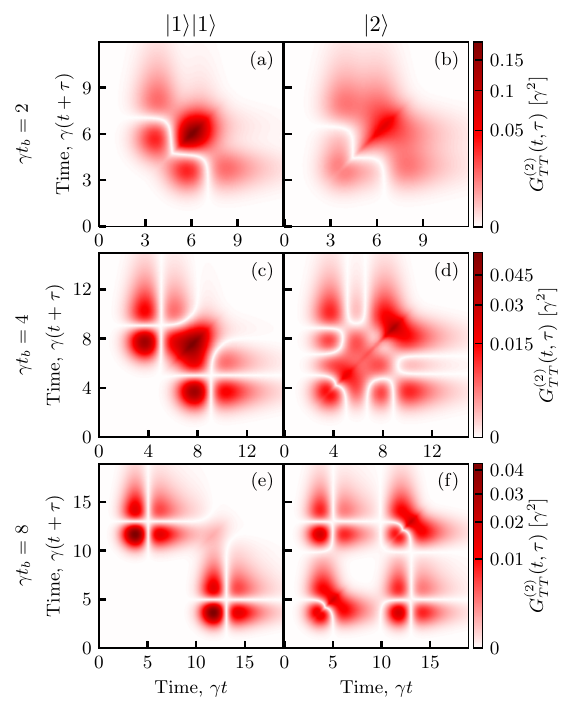}
        \caption{Second-order correlation function $G_{TT}^{(2)}(t,\tau)$ of the same simulations as \Cref{fig:fan_sym_pops}.}
        \label{fig:sym_G2_TT}
    \end{minipage}
    \hfill
    \begin{minipage}{0.495\linewidth}
        \centering
        \includegraphics[width=\linewidth]{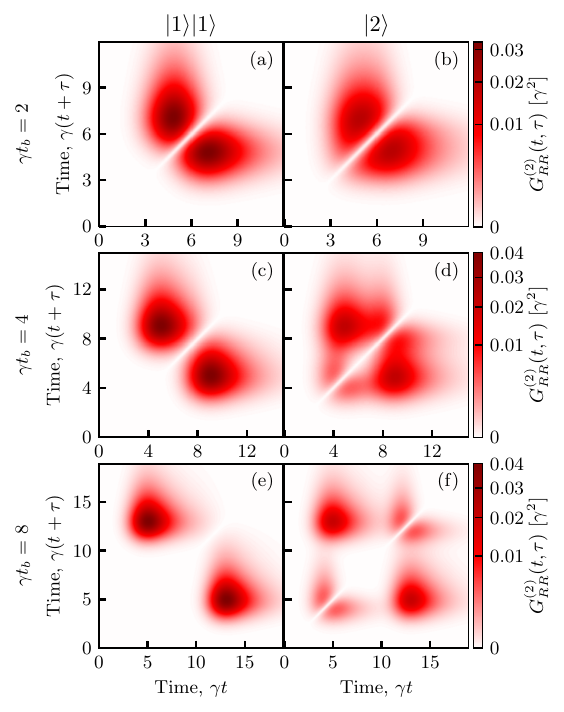}
        \caption{Second-order correlation function $G_{RR}^{(2)}(t,\tau)$ of the same simulations as \Cref{fig:fan_sym_pops}.}
        \label{fig:sym_G2_RR}
    \end{minipage}
    \centering
    \includegraphics[width=0.495\linewidth]{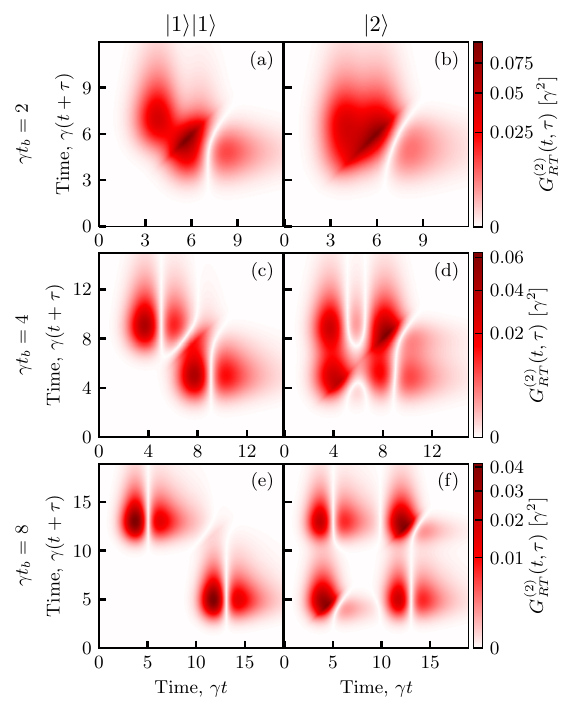}
        \caption{Second-order correlation function $G_{RT}^{(2)}(t,\tau)$ of the same simulations as \Cref{fig:fan_sym_pops}.}
    \label{fig:sym_G2_RT}
\end{figure}

To supplement the results presented in the main text, we have included here the population dynamics (\Cref{fig:fan_sym_pops}) and the second-order correlation functions (\Cref{fig:sym_G2_TT,fig:sym_G2_RR,fig:sym_G2_RT}) resulting from the scattering of the Gaussian input states off of a symmetrically coupled two level system (TLS). 

Since the TLS is coupled symmetrically, in addition to the transmitted second-order correlation function, $G_{TT}^{(2)}(t,\tau)$, we can also calculate the reflected 
second-order correlation function, $G_{RR}^{(2)}(t,\tau)$, and the mixed second order correlation function, $G_{RT}^{(2)}(t,\tau)$. We  note that for a desired effect of strong transmitted correlations in the field, the symmetric coupling causes an alteration in the structure of this correlation and greatly reduces the magnitude of the transmitted correlations due to the backscattering of photons resulting in probabilities of doubly reflected photons, or one photon reflected while the other is transmitted.

\section{Matrix Product Factorization of Fock States}
\label{app:mps_fock_states}

In this section, we discuss the matrix product factorization of Fock states used for the 2-photon case to yield the tensors presented in the main text.
We also elaborate on the complexity scaling of this method for Fock states with general pulse envelopes.

This analysis is based on the ideas presented in Ref.~\cite{bacon2008wfa}, which approaches the creation of the matrices from the perspective of formal language theory over the alphabet of the number of photons that might be placed in each individual time bin. The main idea of the process is, since we are working in the Fock state basis, to use the bond indices to track the number of photons that have been placed in the MPS chain up to a given point, analogous to the memory of a finite automaton. In this way, we think of the MPS as encoding the complex coefficients associated with a given sequence of photon numbers in bins, with the overall state being the superposition of the sum of these correctly weighted states.

We begin by considering a 2-photon Fock state with an arbitrary wavefunction propagating in a single channel, given by
\begin{align}
    \ket{\psi_{\rm in}} =& \frac{1}{\sqrt{2!}}\left( \iint dt_1 dt_2 f(t_1,t_2) a^\dag(t_1) a^\dag(t_2) \right)\ket{0} \nonumber\\
    \overset{\rm discretize}{\rightarrow}=& \frac{1}{\sqrt{2!(\Delta t)^2}} \left(\sum_{j=1}^m \sum_{k=1}^m f(t_j,t_k)\Delta B^\dag(t_j) \Delta B^\dag(t_k) \right)\ket{0}. 
    \label{eq:discretized_fock}
\end{align}

In the case of a product state of $N$ indistinguishable photons, the derivation of the matrices is presented in Ref.~\cite{FanMPSComparison}, and results in matrices with a bond dimension of $N+1$ that simply track the placed number of photons from 0 to $N$. This simple result is made possible by a multinomial expansion that is not feasible when the product states have different temporal envelopes, or for more complicated states.

In the case of a general 2-photon product state, we instead get
\begin{align}
    \ket{\psi_{\rm in}} =& \frac{1}{\sqrt{2!(\Delta t)^2}} \left(\sum_{j=1}^m \sum_{k=1}^m f^{(1)}(t_j)f^{(2)}(t_k)\Delta B^\dag(t_j) \Delta B^\dag(t_k) \right)\ket{0} \nonumber\\
    =& \frac{1}{\sqrt{2!}}\left( \sum_{j=1}^m \sqrt{2!}f_j^{(1)}f_j^{(2)} \ket{2_j} + \sum_{j=1}^m \sum_{k=1}^m f_j^{(1)}f_k^{(2)} \ket{1_j,1_k}\right),
\end{align}
where $f_j^{(k)} = \frac{f^{(k)}(t_j)}{\sqrt{\Delta t}}$. Each individual state in the above superposition can be compactly described in a single nondeterministic complex weighted finite automaton with time bin-dependent transitions. The overall state is given by the sum over all possible transitions to an accepting/final state in the automaton.

Formally, such an automaton is defined as a 5-tuple, $(Q,\Sigma,W,\alpha,\Omega)$, where $Q$ is a finite set of states, $\Sigma$ is the alphabet of the automaton, $W$ is the time dependent weight function that maps transitions between states on some output letter to a complex number, that is $W:Q\times\Sigma\times Q\times\N \rightarrow \C$, $\alpha$ is the complex-valued initial distribution and has a shape $1\times Q$, and $\Omega$ is the complex-valued final distribution with shape $Q\times 1$ \cite{bacon2008wfa}. A standard state diagram is shown as an example of such an automaton for the 2-photon product state in \Cref{fig:automata_N=2_product}. In this figure, the states, $Q$, are depicted in circles around their labels and track the photons placed from each temporal envelope. The alphabet, $\Sigma=\{0,1,2\}$, describes the occupation of a photon (the eigenvalues of the number operator for our time bin basis size) and is the first symbol labeling the transitions (before the slash). The second number (following the slash) is the complex output/weight of the transition and is the output of $W$. For the purposes of Fock states, $\alpha$ and $\Omega$ are trivial, as they simply denote 0 placed photons and all $N$ photons placed for an $N$-photon state. This means that the initial state (labeled by an arrow into the diagram) has a weight of 1 in $\alpha$, and the final state, denoted by the double circles, has a weight of 1 in $\Omega$, while all other elements of the initial and final distributions are 0. This representation acts as a helpful interpretation of the state, as it directly describes a matrix product state factorization, with $|Q|$ being the size of the bond dimension of the matrix product state, and $W$ describing the $|\Sigma|$ matrices at each site, $k$, in the chain ($\alpha$ and $\Omega$ can be contracted with the first and last matrices in the chain to yield the forms presented in the main text).

\begin{figure}[h]
    \centering
    \input{paper_figs/tikz/automata\_N=2.tikz}
    \caption{A complex weighted automaton for a 2-photon product Fock state. The states (denoted in circles) track the number of photons and which temporal envelope they are from. The arrows dictate transitions between states, the first number along each transition denotes the number of photons at the given bin ($k$), and the quantity after the slash denotes the factor contributed to the state from the given transition.
    }
    \label{fig:automata_N=2_product}
\end{figure}

From the automata, the matrices used for MPS can directly be created based on the transitions, states, and transition weights, with the number of states ($|Q|$) being equal to the bond dimension of the matrices \cite{bacon2008wfa}. In general, for an $N$-photon product state a bond dimension of at most $2^N$ is necessary to construct the tree that tracks each possible ordering in which the $N$ photons might be placed and results from a subset construction of the labelled temporal modes, $\sum_{k=0}^N \binom{N}{k} = 2^N$.

While MPS easily extends to higher numbers of quanta, the strength of the scattering theory lies in the easy accessibility of the full scattered wavefunction and its capacity to handle arbitrary input 2-photon states. To consider an arbitrary 2-photon state with MPS, we must first conduct a Schmidt decomposition on the function, $f(t_1,t_2)$, expressing it as the sum of a product of single-time-point functions,
\begin{align}
    f(t_1,t_2) \approx \sum_{\beta=1}^{\chi_{\rm max}} s_\beta A^{(1)}_\beta(t_1)A^{(2)}_\beta(t_2),
\end{align}
where $\chi_{\rm max}$ is some finite cutoff of the basis functions determined by the entanglement of the given state and $s_\beta$ are the Schmidt coefficients. Using this, an arbitrary 2-photon state can be written as
\begin{align}
    \ket{\psi_{\rm in}} =& \frac{1}{\sqrt{2!(\Delta t)^2}} \left(\sum_{j=1}^m \sum_{k=1}^m \sum_{\beta=1}^{\chi_{\rm max}} s_\beta A^{(1)}_\beta(t_j)A^{(2)}_\beta (t_k)\Delta B_j^\dag \Delta B_k^\dag \right)\ket{0} \nonumber\\ 
    =& \frac{1}{\sqrt{2!}} \sum_{\beta=1}^{\chi_{\rm max}} \Bigg( \sum_{j=1}^m \sqrt{2!}A_\beta^{(1)}(j) A_\beta^{(2)}(k) \ket{2_j} + \sum_{j=1}^m \sum_{k=1}^m A_\beta^{(1)}(j) A_\beta^{(2)}(k) \ket{1_j,1_k}\Bigg),
\end{align}
where we let $A_\beta^{(l)}(k) \equiv \frac{\sqrt{s_\beta}}{\sqrt{\Delta t}}A_\beta^{(l)}(t_k)$. This is simply a sum over $\chi_{\rm max}$ product states, but each mode of the Schmidt decomposition of the two time point temporal pulse envelope must be tracked separately. As such, its associated weighted finite automaton will require $2+2\chi_{\rm max}$ states to track which mode, $\beta$, the single photon contributions are from, and so too will the bond dimension scale with this complexity.

While the complexity of the bond dimension with respect to $\chi_{\rm max}$ remains tame for small $N$, being linear in the case that $N=2$, as $N$ increases, Fock states with large entanglement quickly become intractable. In general, larger arbitrary $N$ point functions require iterative Schmidt decompositions to reduce them to a $N-1$ sums over the products of single-time-point functions,
\begin{align}
    f(t_1,t_2,\dots,t_N) \approx& \sum_{\beta_1,\beta_2,\dots,\beta_{N-1}}^{\chi_{\rm max}} A_{\beta_1}^{(1)}(t_1) A_{\beta_1\beta_2}^{(2)}(t_2)\dots A_{\beta_{N-2}\beta_{N-1}}^{(N-1)}(t_{N-1})A_{\beta_{N-1}}^{(N)}(t_{N}), \label{eq:schmidtDecomp}
\end{align}
where $\chi_{\rm max}$ is some truncation of the maximum Schmidt labels that determines the numerical precision of the approximation. 
Tracking these modes results in a combinatorial explosion, as one finds that the number of states necessary to track the different ways to place $k$ photons scales as $O(N^k\chi_{\rm max}^{2k})$, with the bond dimension of the MPS factorization necessitating the sum over the number of states needed to track $0$ to $N-1$ photons.

In general, the bond dimension for the input state can be written exactly in the form
\begin{align}
    |Q|(N,\chi_{\rm max}) = 2 + \sum_{k=1}^{N-1} |Q|_k(N,\chi_{\rm max}),
\end{align}
where the first term accounts for the initial and final states, and $|Q|_k(N,\chi_{\rm max})$ is the number of states needed to track the different possibilities of placing $k$ photons over the various Schmidt modes, $A^{(j)}(t_j)$. 

The function, $|Q|_k(N,\chi_{\rm max})$, can be determined explicitly by summing over the number of different adjacent Schmidt modes in \Cref{eq:schmidtDecomp}, which determines the number of free Schmidt indices. Additionally, it is important to note that the end tensors in \Cref{eq:schmidtDecomp} have only a single free Schmidt index, and the "placement" of a photon in such a mode changes the number of free Schmidt indices. Noting these, we can sum over the number of possible adjacent groups (which is limited by the number of photons, $k$) and the number of edge modes included (a maximum of 2) to yield the equation,
\begin{align}
    |Q|_k(N,\chi_{\rm max}) = \sum_{a=1}^{k}\sum_{e=0}^{\min(a,2)} C_k^{\rm (e)}(N,a)  \chi_{\rm max}^{k+a-e}
\end{align}
where $a$ counts the number of adjacency groups of Schmidt modes included in the given state, $e$ counts the number of end Schmidt modes included in the state, and the $\chi_{\rm max}^{k+a-e}$ factor is due to the $\chi_{\rm max}$ different possible values for each free Schmidt index, which is reduced by the inclusion of the end modes or many adjacent modes (for example, a photon in mode $A_{\beta_1}^{(1)}$ fixes the index $\beta_1$ for a photon in mode $A_{\beta_1\beta_2}^{(2)}$).

To determine the coefficients $C_k^{(e)}(N,a)$, we need to count the number of possible ways to distribute these $a$ adjacency groups of $k$ Schmidt modes out of a total of $N$ modes, of which $e$ of the edge modes are selected. This can be understood as being an equivalent problem to determining the number of binary strings of length $N$ with $k$ ones and $a$ groups of contiguous ones, with end conditions of the string dependent on $e$. Considering first $C_k^{(0)}(N,a)$, we note that partitioning some $k$ chosen modes into $a$ different adjacency groups can be done in $\binom{k-1}{a-1}$ different ways, as this is determining the number of ways to place $a-1$ bars in the $k-1$ spaces between the $k$ selected modes to get $a$ different contiguous groups of modes. Each of these contiguous groups must be separated by at least one Schmidt mode of the decomposition, reducing the number of unselected modes to choose by $a-1$. Additionally, if $e=0$, the two end modes must not be placed, reducing the modes to select by $a+1$ and meaning there are only $N-k-a-1$ unselected modes that can be freely placed in the spaces between the adjacency groups. A stars and bars argument can be used to determine that the number of combinations of ways to place these unselected modes in the $a$ gaps between adjacency groups is then given by $\binom{(N-k-a-1)-(-a)}{a}=\binom{N-k-1}{a}$. The argument for $e\neq0$ follows similarly, with the first factor being the same and the second factor having $e$ fewer choices as $e$ adjacency groups are fixed in location due to the inclusion of the end sites (but this also provides an additional unoccupied mode to freely place). Thus we have
\begin{align}
    C_k^{\rm (0)}(N,a) =& \binom{k-1}{a-1}\binom{N-k-1}{a} \\
    C_k^{\rm (1)}(N,a) =& 2\binom{k-1}{a-1}\binom{N-k-1}{a-1} \\
    C_k^{\rm (2)}(N,a) =& \binom{k-1}{a-1}\binom{N-k-1}{a-2},
\end{align}
where the factor of 2 for the second case comes from the symmetry of the left/right end occupation of the Schmidt modes of $f$.

In the case that $\chi_{\rm max}=1$ by repeated applications of Pascal's identity, extending the sum over $a$, and using the Chu-Vandermonde identity we can show that $|Q|_k(N,1) = \binom{N}{k}$, in agreement with the expected result when the input pulse envelope, $f$, has a Schmidt rank of 1.

\section{Analytical Populations for the Two-Level Atom}

Although we have computed various observables in the main text using MPS and scattering theory, there is potential merit also in obtaining some of the observables analytically, as they may yield additional insight.
Below, we show how to do that, but generally, we find that the final form of the equations is not so useful for any extra insight beyond what can be computed with MPS and scattering theory. However, we include them for completeness.

We can determine the dynamics for product state input Fock states using a hierarchy of equations approach, similar to the one discussed in Ref.~\cite{PhysRevA.90.063832}. We begin by defining the functional input field creation operators for a pulse with bandwidth:
\begin{align}
    A_{\rm in}^{\mu\dag}[g] = \int dt g(t)a_{\rm in}^{\mu\dag},
\end{align}
which satisfy the commutation relations
\begin{align}
    [a_{\rm in}^\mu(t), A_{\rm in}^{\eta\dag}[g]] =& \delta_{\mu\eta}g(t) \\
    [A_{\rm in}^\mu[g_1], A_{\rm in}^{\eta\dag}[g_2]] =& \delta_{\mu\eta} \int dt g_1^*(t)g_2(t).
\end{align}

We then have the input field state given by
\begin{align}
    \ket{\psi_{\rm in}} = \frac{K}{\sqrt{2}} \sum_{i,j=1}^2 (c_1 c_2^T)_{ij} A_{\rm in}^{\rm{R} \dag}[g_i]A_{\rm in}^{\rm{R} \dag}[g_j]\ket{0},
\end{align}
for the appropriate normalization constant $K$ and with the coefficients
\begin{align}
    c_1 = \begin{pmatrix}
        1-\alpha \\ \alpha
    \end{pmatrix}&&
    c_2 = \begin{pmatrix}
        \alpha \\ 1-\alpha
    \end{pmatrix}&&
    C = c_1 c_2^T = \begin{pmatrix}
        \alpha(1-\alpha) & (1-\alpha)^2 \\
        \alpha^2 & \alpha(1-\alpha)
    \end{pmatrix}.
\end{align}

From input/output theory \cite{PhysRevA.31.3761}, we 
can derive the operator equations of motion
\begin{align}
    a_{\rm out}^\mu(t) =& a_{\rm in}^\mu(t) - i\sqrt{\gamma_\mu}\sigma^-(t)\nonumber\\
    \frac{d\sigma^-(t)}{dt} =& -\left(i\Delta + \frac{\Gamma}{2}\right)\sigma^-(t) + i\sum_{\mu=\rm L,R} \sqrt{\gamma_\mu} \sigma^z(t)a_{\rm in}^\mu(t) \nonumber\\
    \frac{d ( \sigma^+ (t) \sigma^-(t)) }{dt} =& -\Gamma\sigma^+ (t)\sigma^-(t) - \sum_{\mu=\rm L,R} i\sqrt{\gamma_\mu} \big[ \sigma^+ (t) a^\mu_{\rm in}(t)- a_{\rm in}^{\mu \ \dag}(t) \sigma^- (t)\big],
\end{align}
where $\Gamma\equiv \sum_\mu \gamma_\mu$ is the sum of the decay rates into the different channels.

From these equations of motion, under the initial conditions that the TLS population and coherence are 0 at time $t_0=-\infty$, we have
\begin{align}
    P^{(0)}(t) &= 0 \nonumber\\
    C_{i}^{(1)}(t) &= -i\sqrt{\gamma_{\rm R}}\int_{-\infty}^t ds e^{-\zeta(t-s)}g_i(s)\\
    P_{i,j}^{(1)}(t) &= i\sqrt{\gamma_{\rm R}}\int_{-\infty}^t ds e^{-\Gamma(t-s)}\left( g_i^*(s)C_{j}^{(1)}(s) - \left(g_j^*(s)C_{i}^{(1)}(s)\right)^*\right) \\
    C_{i,jk}^{(2)}(t) &= i\sqrt{\gamma_{\rm R}}\int_{-\infty}^t ds e^{-\zeta(t-s)}\left( g_j(s)(2P_{i,k}^{(1)}(s) - \braket{1_{g_i}^{\rm R}|1_{g_k}^{\rm R}}) + g_k(s)(2P_{i,j}^{(1)}(s) - \braket{1_{g_i}^{\rm R}|1_{g_j}^{\rm R}})\right) \\
    P_{ij,kl}^{(2)}(t) &= i\sqrt{\gamma_{\rm R}}\int_{-\infty}^t ds e^{-\Gamma(t-s)}\left[ g_j^*(s)C_{i,kl}^{(2)}(s) + g_i^*(s)C_{j,kl}^{(2)}(s) - \left(g_k^*(s)C_{l,ij}^{(2)}(s) + g_l^*(s)C_{k,ij}^{(2)}(s)\right)^*\right],
\end{align}
where we have defined $\zeta\equiv i\Delta + \frac{\Gamma}{2}$, and have
\begin{align}
    P^{(0)}(t) &= \braket{0|\sigma^+\sigma^-|0}(t) \\
    C^{(1)}_i(t) &= \braket{0|\sigma^-|1_{g_i}^{\rm R}}(t) = \braket{0|\sigma^- A_{\rm in}^{\rm R \dag}|0}(t) \\
    P^{(1)}_{i,j}(t) &= \braket{1_{g_i}^{\rm R}|\sigma^+\sigma^-|1_{g_j}^{\rm R}}(t) = \braket{0| A_{\rm in}^{\rm R}[g_i] \sigma^+\sigma^- A_{\rm in}^{\rm R \dag}[g_j]|0}(t) \\
    C^{(2)}_{i,jk}(t) &= \braket{1_{g_i}^{\rm R}|\sigma^-|1_{g_j}^{\rm R}1_{g_k}^{\rm R}}(t) = \braket{0| A_{\rm in}^{\rm R}[g_i] \sigma^- A_{\rm in}^{\rm R \dag}[g_j]A_{\rm in}^{\rm R \dag}[g_k]|0}(t) \\
    P^{(2)}_{ij,kl}(t) &= \braket{1_{g_i}^{\rm R}1_{g_j}^{\rm R}|\sigma^+\sigma^-|1_{g_k}^{\rm R}1_{g_l}^{\rm R}}(t) = \braket{0| A_{\rm in}^{\rm R}[g_i]A_{\rm in}^{\rm R}[g_j] \sigma^+\sigma^- A_{\rm in}^{\rm R \dag}[g_k]A_{\rm in}^{\rm R \dag}[g_l]|0}(t). 
\end{align}

By defining the pulse envelope filtered by the TLS coherence kernel, we can let
\begin{align}
    F_i(t) = \int_{-\infty}^t ds e^{-\zeta(t-s)} g_i(s),
\end{align}
and with that concisely write the population in the single photon sector for the early equations in the hierarchy as
\begin{align}
    C_{i}^{(1)}(t) &= -i\sqrt{\gamma_{\rm R}}F_i(t)\\
    P_{i,j}^{(1)}(t) &= \gamma_{\rm R}\int_{-\infty}^t ds e^{-\Gamma(t-s)}\left( g_i^*(s)F_{j}(s) + \left(g_j^*(s)F_{i}(s)\right)^*\right).
\end{align}

We then note that we can decompose the coherence in the 2-photon sector into a term dependent on the population in the single photon sector, and a term dependent only on pulse activity
\begin{align}
    C_{i,jk}^{(2)}(t) = C_{i,jk}^{(\rm pop)}(t) + C_{i,jk}^{(\rm pulse)}(t),
\end{align}
where
\begin{align}
    C_{i,jk}^{(\rm pop)}(t) &= 2i\sqrt{\gamma_{\rm R}}\int_{-\infty}^t ds e^{-\zeta(t-s)}\left( g_j(s)P_{i,k}^{(1)}(s) + g_k(s)P_{i,j}^{(1)}(s)\right) \\
    C_{i,jk}^{(\rm pulse)}(t) &= -i\sqrt{\gamma_{\rm R}}\int_{-\infty}^t ds e^{-\zeta(t-s)}\left( \Lambda_{ik}g_j(s) + \Lambda_{ij}g_k(s)\right) \nonumber\\
    &= -i\sqrt{\gamma_{\rm R}}\left( \Lambda_{ik}F_j(t) + \Lambda_{ij}F_k(t)\right),
\end{align}
where we define $\Lambda_{ij} = \braket{1_{g_i}^{\rm R}|1_{g_j}^{\rm R}} = [A_{\rm in}^{\rm R}[g_1], A_{\rm in}^{\rm R\dag}[g_2]]$.

Using this, we can express the population in the 2-photon sector in terms of these two coherence contributions
\begin{align}
    P_{ij,kl}^{(2)}(t) = P_{ij,kl}^{\rm(sat)}(t) + P_{ij,kl}^{\rm(pulse)}(t),
\end{align}
where
\begin{align}
    P_{ij,kl}^{\rm(sat)}(t) &= i\sqrt{\gamma_{\rm R}}\int_{-\infty}^t ds e^{-\Gamma(t-s)} \left[ g_j^*(s)C_{i,kl}^{\rm(pop)}(s) + g_i^*(s)C_{j,kl}^{\rm(pop)}(s) \right . \nonumber \\ 
    & \left . \mkern210mu  - \left(g_k^*(s)C_{l,ij}^{\rm(pop)}(s) + g_l^*(s)C_{k,ij}^{\rm(pop)}(s)\right)^*\right] \\
    P_{ij,kl}^{\rm(pulse)}(t) &= \gamma_{\rm R}\int_{-\infty}^t ds e^{-\Gamma(t-s)}\Big( g_j^*(s)\left( \Lambda_{il}F_k(s) + \Lambda_{ik}F_l(s)\right) + g_i^*(s)\left( \Lambda_{jl}F_k(s) + \Lambda_{jk}F_l(s)\right) \nonumber\\
    &+ \left(g_k^*(s)\left( \Lambda_{lj}F_i(s) + \Lambda_{li}F_j(s)\right) + g_l^*(s)\left( \Lambda_{kj}F_i(s) + \Lambda_{ki}F_j(s)\right)\right)^*\Big).
\end{align}


The population of the TLS for our input state of interest is given by
\begin{align}
    P_{\rm tot} =& \braket{\psi_{\rm in}| \sigma^+\sigma^-(t)|\psi_{\rm in}} = \frac{|K|^2}{2} C_{ij} C_{kl} \left(P_{ij,kl}^{\rm(sat)}(t) + P_{ij,kl}^{\rm(pulse)}(t)\right),
\end{align}
where we have used Einstein summation convention. Note that, while $P_{ij,kl}^{(2)}$ has 16 elements, only 6 of them are independent due to the permutation symmetry of the index pairs and the Hermiticity of the expectation values. 

This approach can handle arbitrary pulse envelopes, as well as a larger basis of envelopes. Restricting our consideration to real Gaussian pulses in the time domain that are on resonant so that $\Delta=0$ and $\zeta\in\R$, we have that $g_i(t),F_i(t),\Lambda_{ij}\in\R$. For normalized single pulse envelopes we have
\begin{align}
    g_1(t) = \frac{1}{(\pi\sigma^2)^{1/4}}e^{-(t-t_0)^2/2\sigma^2},
\end{align}
and $g_2(t) = g_1(t-t_b) = g(t-t_b)$.

Under these circumstances, we have the filtered pulses
\begin{align}
    F_1(t) = \pi^{1/4}\sqrt{\frac{\sigma}{2}} e^{-\frac{\Gamma}{2}\left(t - t_0\right) + \frac{\Gamma^2\sigma^2}{8}}\erfc\left( \frac{\frac{\Gamma\sigma^2}{2} -(t-t_0)}{\sqrt{2}\sigma}\right),
\end{align}
and since we take the initial time $t_0=-\infty$ we have that $F_2(t) = F_1(t-t_b) = F(t-t_b)$. Lastly, we have the projections
\begin{align}
    \Lambda_{11} =& \Lambda_{22} = 1 \\
    \Lambda_{12}(t_b) =& \Lambda_{21}^*(t_b) = \Lambda_{21}(t_b) = e^{-t_b^2/4\sigma^2} .
\end{align}

We highlight that all of these quantities are nonnegative in the general case of real temporal envelopes on resonance. As a result all $P_{ij,kl}^{\rm (pulse)}$ terms are nonnegative, while the saturation term $P_{ij,kl}^{\rm(sat)}$ terms are all nonpositive and act as a resistive force against TLS saturation in the case of 2-photon excitation, preventing a population greater than 1 that might occur in the case of a harmonic oscillator. These two competing effects both grow larger in magnitude for smaller pulse separations, $t_b$.

With these results we have what is needed to determine the six independent population terms, $P^{(2)}_{11,11},P^{(2)}_{11,12},P^{(2)}_{11,22},P^{(2)}_{12,12},P^{(2)}_{12,22},P^{(2)}_{22,22}$. Considering the terms not dependent on the single photon sector populations, we have
\begin{align}
    P_{11,11}^{\rm(pulse)}(t) &= 8\gamma_{\rm R}\int_{-\infty}^t ds e^{-\Gamma(t-s)}g(s)F(s) \\
    P_{11,12}^{\rm(pulse)}(t) &= 2\gamma_{\rm R}\int_{-\infty}^t ds e^{-\Gamma(t-s)}\Big( g(s) F(s-t_b) + g(s-t_b)F(s) + 2\Lambda_{12}(t_b)g(s)F(s)\Big) \\
    P_{11,22}^{\rm(pulse)}(t) &= 4\gamma_{\rm R}\int_{-\infty}^t ds e^{-\Gamma(t-s)}\Lambda_{12}(t_b)\Big( g(s)F(s-t_b) + g(s-t_b)F(s)\Big) \\
    P_{12,12}^{\rm(pulse)}(t) &= 2\gamma_{\rm R}\int_{-\infty}^t ds e^{-\Gamma(t-s)}\Big( g(s)F(s) + g(s-t_b)F(s-t_b) 
    \nonumber \\
    &+ \Lambda_{12}(t_b)\left( g(s-t_b)F(s) + g(s)F(s-t_b)\right)\Big) \\
    P_{12,22}^{\rm(pulse)}(t) &= 2\gamma_{\rm R}\int_{-\infty}^t ds e^{-\Gamma(t-s)}\Big( g(s)F(s-t_b) + g(s-t_b)F(s) + 2\Lambda_{12}(t_b)g(s-t_b)F(s-t_b)\Big) \\
    P_{22,22}^{\rm(pulse)}(t) &= 8\gamma_{\rm R}\int_{-\infty}^t ds e^{-\Gamma(t-s)} g(s-t_b)F(s-t_b).
\end{align}
Likewise, the $P_{ij,kl}^{(\rm sat)}$ terms can be calculated, and include additional nestings of convolutions. We could also calculate field observables, such as the transmitted flux, using the same formalism.

Overall, although these are in analytic form, they do not offer much in terms of analytical insight to our main simulations. Nevertheless, they are useful, and quite general.

\bibliography{references_all}

%% file: paper_figs/tikz/automata_N=2.tikz
\usetikzlibrary{automata, positioning,calc, quotes}

\begin{tikzpicture}[inner sep=1mm, draw=black, thick, auto,
every initial by arrow/.style = {thick}, node distance = 3cm]

    \node (0) [state, initial, initial text = {}] {$0$};
    \node (2) [state, accepting, right =of 0] {$2$};
    \coordinate(a) at ($0.5*(0)+0.5*(2)$);
    \coordinate(b) at ($(a) + (0,1)$);


    \node (1b) [state, below =of a] {$1_2$};
    \node (1a) [state, above =of a] {$1_1$};

    \path [-stealth, thick]
     (0) edge[loop above] node {$0_k/1$}   (0)
     (1a) edge[loop right] node {$0_k/1$}   (1a)
     (1b) edge[loop right] node {$0_k/1$}   (1b)
     (2) edge[loop above] node {$0_k/1$}   (2)
     (0) edge[left, pos=0.7] node {$1_k/f_k^{(1)}$}   (1a)
     (0) edge[left, pos=0.5] node {$1_k/f_k^{(2)}$}   (1b)
     (1a) edge[pos=0.3] node[right]{$1_k/f_k^{(2)}$}   (2)
     (1b) edge[right] node {$1_k/f_k^{(1)}$}   (2)
     (0) edge[above] node {$2_k/\sqrt{2!}f_k^{(1)}f_k^{(2)}$}   (2)

     ;
    
\end{tikzpicture}